\UseRawInputEncoding
\documentclass{article}
\usepackage{amsmath}
\usepackage{amsfonts}
\numberwithin{equation}{section}

\newcommand{\be}{\begin{equation}}
\newcommand{\ee}{\end{equation}}
\newcommand{\ba}{\begin{eqnarray}}
\newcommand{\ea}{\end{eqnarray}}

\begin{document}
\hoffset=-.4truein\voffset=-0.5truein
\setlength{\textheight}{8.5 in}
\begin{titlepage}
\begin{center}
\vskip 16 mm

{\large \bf Punctures and $p$-spin curves from matrix models III. \\
$D_l$ type and  logarithmic potential}

\vskip .4in
\begin{center}
 { Shinobu Hikami }
\end{center}
Okinawa Institute of Science and Technology Graduate University, 1919-1 Tancha, Okinawa 904-0495, Japan.
e-mail: hikami@oist.jp
 \end{center}
\vskip 8mm       
{\bf Abstract }   The intersection numbers for  $p$ spin curves of the moduli space $\overline{\mathcal{M}}_{g,n}$ are considered for $D_l$ type by a matrix model.  The asymptotic behavior of the large genus $g$ limit and large $p$ limit are derived. The remarkable features of the cases of  $p= \frac{1}{2}, - \frac{1}{2}, -2, -3$  are examined in the Laurent expansion for multiple correlation functions. The strong coupling expansions for the negative $p$ cases are considered.

 \vskip 8mm
 {\bf{Key words}} Random matrix model, Topological field theory, Intersection numbers
\vskip 3mm
 \end{titlepage}

\vskip 3mm
 \section{Introduction}
 
    It is well known that the generating function of the intersection numbers on moduli space of $p$ spin stable curves becomes the $\tau$ function of the
 generalized KdV hierarchies ($p$-reduced KP hierarchies), which are related to a two dimensional gravity \cite{Witten,DVV}. There are now many  studies of  the intersection numbers including Gelfand-Dikii pseudo differential equation \cite{Gelfand,Witten2}.  We have proposed   a method of the calculation for this intersection numbers by  a generalized Kontsevich matrix model, which was derived by the duality and replica method  based on the random matrix theory \cite{BrezinHikami12,BrezinHikami11,BrezinHikami3,BrezinHikami5}.
 For one point function, an expression for single intersection numbers has been derived explicitly.
 
 For the purpose of the extension to  half integer spin $p$, we reformulated this integral expressions by  a new change of variables, and  we have investigated the  intersection numbers of Ramond punctures for half-spin  in previous articles I, II \cite{BrezinHikami1,BrezinHikami2}.
 
 This reformulation enables us to  obtain easily the intersection numbers for integer $p$ (Neveu-Schwarz punctures) for $n$ point functions, which should be consistent with the results obtained by the recursive method \cite{LiuXu,LiuVakilXu} due to Gelfand-Dikii equation.  The evaluation of several marked points in general $p$ and  for genus $g$ was obtained in the recursive
calculations  \cite{LiuVakilXu}, and we show in this article that our method of the Laurent expansion agrees with them for the lower orders, especially  for three point functions.
 
  The  ADE singularities characterized by Dynkin diagrams are important topics in statistical physics. 
 The  $D_l$ singularity is represented  by  the algebraic equation $ y^{l-1} + y x^2 + z^2=0$ ($l\ge 4$).  
 The Coxeter number defined by $p= 2 l - 2$  in  $D_l$ singularity can be interpreted as  a spin $p$. The weight system of $D_l$ is known as $(a,b,c;h) =(2,l-2,l-1; 2(l-1))$.
 The simply laced Lie algebra, $A_l,B_l,C_l,D_l,$ are related to classical Lie groups $SU(l+1),SO(2l+1), SP(l), SO(2l)$, respectively. We have discussed  $A_l,B_l,C_l,D_l$ cases, where HarishChandra theorem can be  applied for 
 the  random matrix models with external sources \cite{BrezinHikami5,BrezinHikami7}. These random matrix models were applied for  the non-orientable surfaces or Klein surfaces.
  Recently the  $D_l$ singularity has been discussed
 for the intersection numbers of one point function \cite{Dubrovin}.  
 
 The open intersection numbers have been discussed based on the logarithmic matrix models \cite{BrezinHikami10,BrezinHikami4,Pandharipande,Bertola,Safnuk,BrezinHikami6,DijkgraafWitten,Buryak,Alexandrov1}, which shows the extension of the intersection theory of Riemann surface to open Riemann surface, i.e. it has boundaries \cite{Pandharipande,DijkgraafWitten}. The boundaries, similar to the $D$ brane,  are represented by the logarithmic terms. 
 The matrix model with a logarithmic potential  for the open intersection theory is written for general $p$ (generalized Airy matrix model with a logarithmic potential) \cite{BrezinHikami10},
 \be\label{log}
 Z = \int dB e^{-\frac{c}{p+1} {\rm tr} B^{p+1} + k {\rm tr} {\rm log} B + {\rm tr} B \Lambda}
 \ee
 where $B$ is a Hermitian  matrix and $k$ is a parameter. The matrix $\Lambda$ is an external source. This model is called as  Kontsevich-Penner model, when  $p=2$. 
 
 The spin $p$ is related to $l$ as $p = 2l-2$ for $D_l$ type singularity.
  We investigated before the case of Lie algebra $so(N), sp(N)$ and found that the correspondent one point function $u(s)$ has a logarithmic terms \cite{BrezinHikami7}, which makes
 a difference from $A_l$ type. These cases were discussed as a manifestation of the feature of non-orientable surface (Klein surface) further in \cite{BrezinHikami4,BrezinHikami5}. 

 In this paper,
 we explicitly show that the intersection numbers of $D_l$ type can be derived from  the matrix model witha logarithmic term.
 By the  Gaussian integral of $x$ for $ y^{l-1} + x^2 y + z^2$,  we have $1/\sqrt{y}$ which turns to be  a logarithmic potential by the exponentiation. This may give simple explanation  of the appearance of the logarithmic term in the matrix model of $D_l$ singularity. In general, an oscillating integrals of $n$-variables  have  asymptotic expansions with logarithmic terms, related to Newton polygon \cite{Varcenko} and it is well known that such expansion is related to the resolution of the singularities. 
 The $D_l$ type is related to a real algebraic curves and the Euler characteristics $\chi$  of a real algebraic curve is obtained in  the case of $p= -1$  in $D_l$ type $(p=2l-2)$ \cite{BrezinHikami7,HZ,Goulden}.
 
  It is known that   $A_5,A_4,A_3$ (p=6,5,4) singularities  correspond to Ashkin-Teller model, 3-state Potts model and Ising model
 respectively \cite{Zamolodchikov1,Zamolodchikov2}. The  $A_{p-1}$ singularity has a central charge  $C = 2 - 6/p$.  The central charge $C=\frac{1}{2}$ and critical exponent of the energy $\nu$  are consistent with the values of Ising model. The anomalous conformal  dimensions are $\Delta_\epsilon= \frac{4}{p}$, $\Delta_\phi= \frac{p-3}{p(p-2)}$. These dimensions and central charges agree with the well-known values of Ising ($p=4$) and 3-state Potts model ($p=5$). 

 For $D_l$ type singularities, the central charge is also given by $C=2-\frac{6}{p}$, where $p$ is Coxeter number $p= 2l - 2$. The interesting applications are found in the condensed matter physics for the topological excitation of electron at the 
 boundaries, as Majorana fermions \cite{DijkgraafWitten}, and the edge excitation on the boundary as Quantum Hall effect, for instance. Furthermore, in  $D_4$ singularity, there are intriguing Ramond sector \cite{Zamolodchikov3,Fan}, which is related to  the vanishing relation for one point function of $g= 2+ 3m$, $m\in \mathbb{Z}$, for which we will discuss in this paper.
 
 The spin $p$ takes a value of positive integer according to the singularity theory. However, as discussed for $p=-1$, Euler characteristic is obtained from the continuation from positive integer to $p=-1$ in the expression of the intersection numbers. The cases of $p= \frac{1}{2}$ and $p=\frac{3}{2}$, which are fractional spins, have been discussed in our previous papers \cite{BrezinHikami1,BrezinHikami2}. These cases correspond to "fermion", and the genus expansion of one point function agree with the selection rule
 due to Riemann-Roch theorem. As a conformal field theory (CFT) in  two dimensions, $p= \frac{3}{2}$ case
 exists as $\beta\gamma$ system in the supersymmetric non-linear sigma model, which also corresponds to spanning forrest model with a central charge $C=-2$ \cite{Lesage}.  There is an interesting observation that for $p = \frac{1}{2}$ case, the tautological relations become simple \cite{Shadrin,Faber,Shadrin3}. In this paper, by the results of I,II \cite{BrezinHikami1,BrezinHikami2}, we investigate the $m$ point correlation functions of $p=\frac{1}{2}$, and find that the punctures of Ramond type appear in a pairwise as same as $p=-2$ case.  
 
 The case $p=-2$ corresponds to the unitary matrix model of the lattice gauge theory \cite{BG,GW,Mironov,HikamiMaskawa}. The strong coupling region of this case has a character expansion \cite{Samuel, BrezinHikami8}. We consider this character
 expansion \cite{BrezinHikami8} for the negative $p$ case, $p=-2,-3,...$ by the $n$-point function of $U(s_1,...,s_n)= <\prod_i {\rm tr} e^{s_i M}>$, which can be interpreted as Wilson loops. 

 This article is organized as following:   One point intersection numbers of $p$ spin curves for genus $g$  is shown in section 2 for $A_l$.  In section 3, 
 one point functions  of $D_l$ type is evaluated up to $g=11$. In section 4, one point functions  for non-integer $p$ case of $A$ type and $D$ type  are discussed for $p=\frac{3}{2}, \frac{1}{2}, -\frac{1}{2}, -1, -2$. In section 5, the large $g$ and large $p$ limits are discussed. The integrality of the intersection number is discussed in the relation to Bernoulli numbers.
 It is shown that the denominators of the intersection numbers and Bernoulli numbers are  same. In the limit $p\to \infty$, the intersection numbers reduce to Bernoulli numbers \cite{BrezinHikami13}, which are intriguingly connected to  homotopy, differential topology and number theories \cite{Mazur,Milnor,Zagier,Brieskorn}.
 In section 6, the intersection numbers for multi marked points are evaluated, which is consistent with the results by the recursion relations.
The section 7 is devoted to the evaluations of half spin $p=\frac{1}{2}, - \frac{1}{2}$ and the negative integers $p=-2$ and $p=-3$ cases.
  For the negative integer case, the strong coupling expansion  is investigated in
 the relation to the characters of $U(N)$.  In the section 8, we give summary and discussions.

  \vskip 3mm
 \section{One point function for $A_l$ type}
 
 Since the $p$ spin curves of the moduli space has a correspondence to $A_{p-1}$ singularity by mirror symmetry, we use the terminology of $A_l$ type for
 the $l= p-1$ spin curves, which distinguishes the case $D$ type. 
   
  For $A_l$ case, the generating function of the intersection number for one marked point  is expressed as  \cite{BrezinHikami11,BrezinHikami3}
  \ba\label{u}
  u(s) &=& <{\rm tr} e^{s B}> \nonumber\\
  &=&  \frac{1}{ s} \int \frac{du}{2i\pi} e^{-\frac{c}{p+1}[(u+ \frac{s}{2})^{p+1}- (u -\frac{s}{2})^{p+1}]}
    \ea 
  The generating function for the intersection numbers of $A_l$ ($p=l+1$) is evaluated for 
   small $s$ by the replacement $u= (\frac{t}{cs})^{1/p}$,
  \ba\label{uA}
  u(s) &=& \frac{1}{ s p \pi}\cdot \frac{1}{(c s)^{1/p}} \int_0^\infty dt t^{\frac{1}{p}-1} e^{-t}\nonumber\\
  &\times& e^{-\frac{p(p-1)}{3! 4} s^{2+ \frac{2}{p}} c^{\frac{2}{p} } t^{1-\frac{2}{p}} - \frac{p(p-1)(p-2)(p-3)}{5! 4^2} s^{4+\frac{4}{p}}c^{\frac{4}{p}}t^{1-\frac{4}{p}}+ \cdots}
  \ea
  This leads to
  \ba\label{us}
  u(s) &=& \frac{1}{c^{\frac{1}{p}}s^{1+\frac{1}{p}} \pi} [ \Gamma(1+ \frac{1}{p}) - \frac{p-1}{24} z \Gamma( 1 - \frac{1}{p})\nonumber\\
  && + \frac{(p-1)(2p+1)(p-3)}{5760} z^2 \Gamma(1 - \frac{3}{p})
   + \cdots]
  \ea
  where $z= c^{\frac{2}{p}}s^{2+ \frac{2}{p}}$.
  Writing this expansion with the intersection numbers $<\tau_{n,j}>_g$ ($n$ is integer, and spin component $j=0,1,2,...,p-2$) as
  \be\label{cs}
  u(s) = \sum_g <\tau_{n,j}>_g \frac{1}{\pi}\Gamma(1 - \frac{j+1}{p}) c^{\frac{2g-1}{p}} p^{g-1} s^{(2g-1)(1+ \frac{1}{p})},
  \ee
   one obtain the intersection numbers as a polynomial of $p$.
  We have a relation of $n$ and $j$ for non-vanishing intersection numbers,
  \be
  (2g-1)(1+\frac{1}{p}) = n + \frac{j+1}{p}
  \ee
  which comes from the Riemann-Roch relation (RR) for $\tilde s$ marked points by $\tilde s=1$
  \be\label{RR}
  3g-3 + \tilde s=\sum_{i=1}^{\tilde s} n_i + (g-1)(1-\frac{2}{p}) + \frac{1}{p}\sum_{i=1}^{\tilde s} j_i
  \ee
  
  Thus the $s$ dependence appears for one point case ($\tilde s=1$) as a power  $s^{(2g-1)(1+ \frac{1}{p})} = s^{n + \frac{j+1}{p}}$ in (\ref{cs}).
  The intersection numbers ( in the case of $p=2$), are given by the first Chern class $c_1$ or $\psi$, as
  \be\label{Chern}
  <\tau_{n_1} \cdots \tau_{n_s}>_g = \int_{\bar {\mathcal M}_{g,s}} \psi^{n_1}\cdots \psi^{n_s}
  \ee
  where $\bar {\mathcal M}_{g,s}$ is a compactified moduli space with $s$ marked points on genus $g$ Riemann surface.
  The intersection numbers of one point $\tau_{A_l}(g)$ (Coxeter number p= l+1) is thus given as \cite{BrezinHikami1,BrezinHikami2} 
  \ba\label{A}
  <\tau_{n,j}>_{g=1} &=& \frac{p-1}{24}\nonumber\\
  <\tau_{n,j}>_{g=2} &=& \frac{(p-1)(2p+1)(p-3)}{p\cdot 5!\cdot 4^2\cdot 3} \frac{\Gamma(1- \frac{3}{p})}{\Gamma(1 - \frac{1+j}{p})}\nonumber\\
  <\tau_{n,j}>_{g=3}  &= & \frac{(p-1)(2p+1)(p-5)(8p^2-13p-13)}{p^2\cdot 7!\cdot 4^3\cdot 3^2} \frac{\Gamma(1- \frac{5}{p})}{\Gamma(1- \frac{1+j}{p})}\nonumber\\
  <\tau_{n,j}>_{g=4} &=& \frac{(p-1)(2p+1)(p-7)(72p^4-298p^3-17 p^2 + 562 p + 281)}{p^3\cdot 9!4^4\cdot 15}\nonumber\\
  && \times \frac{\Gamma(1 - \frac{7}{p})}{\Gamma(1 - \frac{1+j}{p})}\nonumber\\
  <\tau_{n,j}>_{g=5} &=& (p-1)(2p+1)(p-3)(p-9)(4p+3)(32 p^4-162 p^3 + p^2\nonumber\\
  && + 326 p + 163)\frac{1}{p^4 11!4^5 3} \frac{\Gamma(1-\frac{9}{p})}{\Gamma(1- \frac{1+j}{p})}\nonumber\\
  <\tau_{n,j}>_{g=6} &=& (p - 1) ( 2 p + 1)(p - 11) ( 530688 p^8 - 5830544 p^7 + 16589332 p^6\nonumber\\
  && + 8955300 p^5 - 65056373 p^4 - 26944928 p^3 + 85178190 p^2\nonumber\\
  && + 80708428 p +  20177107) \frac{1}{p^5\cdot 13! 7\cdot 5\cdot 4^6 3^3}\frac{\Gamma(1- \frac{11}{p})}{\Gamma( 1- \frac{1+ m}{p})}
  \ea
  where $<\tau_{n,j}>_{g=1}= <\tau_{1,0}>_{g=1}$ \cite{BrezinHikami3}. 
  In \cite{BrezinHikami5}, $<\tau_{n,j}>$ up to $g=9$ is evalated. 
 
   When $p=2$ (Kontsevich model), it leads to a simple expression,
  \be\label{Kontsevich}
  <\tau_{3g-2}> = \frac{1}{g! (24)^g}
  \ee
  
  There appear interesting  vanishing relations for $<\tau_{A_l}>$. For instance,ath
  $p=3$ case, $<\tau_{A_l}>$ are vanishing at $g=2+ 3m$ ($m\in \mathbb{Z}$). For $p=5$, $<\tau_{A_l}>$ are vanishing at $g=3+ 5m$ ($m\in \mathbb{Z}$).
  In general odd integer $p$, $<\tau_{A_l}>$ are vanishing at $g= (p+1)/2 + pm$ ($m\in \mathbb{Z}$). Some of these relations can be seen in the expressions of $<\tau_{A_l}>$ up to genus $g=9$ in \cite{BrezinHikami5}. We will see later that $D_4$ ($p=6$) type  has this periodicity of the vanishing relation  at $g= 2+ 3m$ ($m\in \mathbb{Z}$) for $D_4$.   
  
  For the case $A_{2}$ ($p=3$), one point function $u(s)$ is given by the Airy function as \cite{BrezinHikami3}
  \ba\label{Airy}
  u(s) &=& \frac{1}{Ns(3Ns)^{1/3}}A_i(\zeta)\nonumber\\
  &=& \frac{1}{Ns(3Ns)^{1/3}} \biggl[ A_i(0) \biggl(1 + \frac{1}{3!}\zeta^3 + \frac{1\cdot 4}{6!} \zeta^6 + \frac{1\cdot 4\cdot 7}{9!}\zeta^9 + \cdots \biggr)\nonumber\\
  && + A_i'(0) \biggl(\zeta + \frac{2}{4!}\zeta^4 + \frac{2\cdot 5}{7!}\zeta^7 + \frac{2\cdot 5\cdot 8}{10!}\zeta^{10}+\cdots \biggr)\biggr]
  \ea
  where $\zeta =-N^{2/3}(4\cdot 3^{1/3})^{-1} s^{8/3}$, $A_i(0)= 3^{-2/3}/\Gamma(\frac{2}{3})$ and $A_i'(0) = - 3^{-1/3}/\Gamma(\frac{2}{3})$.
  This Airy function leads to the intersection numbers of 
  \be
  <\tau_{\frac{8g-5-j}{3},j}>_g = \frac{1}{(12)^gg!} \frac{\Gamma(\frac{g+1}{3})}{\Gamma(\frac{2-j}{3})}
  \ee
  which shows the vanishing relations for $g=2,5,8,...$ ($g=2+3m$, $m\in \mathbb{Z}$), for such case the value of spin $j$ takes 2.
  The absence of $g=2,5,8,...$ is due to Stokes phenomena.
  
  For $p=4$,  $u(s)$ is written by the Bessel function, \cite{BrezinHikami5}
  \ba
  u(s) &=& \frac{1}{2\sqrt{8}} e^{\frac{3}{160}s^5}\frac{1}{2{\rm sin}(\frac{\pi}{4})}\biggl[ I_{-\frac{1}{4}}(\frac{1}{32}s^5) + I_{\frac{1}{4}}(\frac{1}{32}s^5)\biggr] \nonumber\\
  &=& \frac{1}{8}\sum_{m,n=0}^\infty \frac{1}{m!n!\Gamma(n+\frac{5}{4})}(\frac{3}{160})^m (\frac{1}{64})^{2n + \frac{1}{4}} s^{5m+ 10 n + \frac{1}{4}}
  \ea
  
  We have  Riemann-Roch relation of (\ref{RR})
 for the s-point intersection numbers  $<\tau_{n_1,j_1} \tau_{n_2,j_2} \cdots \tau_{n_s,j_s}>_g$. 
 The factor $(1- \frac{2}{p})$ is a central charge $\hat c = \frac{p-2}{p}$. The last term is also charge for $p$ spin curves. 
  Note this central charge $\hat c$ is also valid for $D_l$ singularity, since the weight system of $D_l$ singularity $W = y^{l-1} + y x^2$, $q_y= \frac{1}{l-1}$,
  $q_a = \frac{l-2}{2(l-1)}$. The central charge $\hat c$ is given by $ 2 - 2 q_y - 2 q_x= 1- \frac{1}{l-1}$. Since $p= 2(l-1)$, we have $\hat c = 1-\frac{2}{p}$
  for $D_l$ case. Thus the Riemann-Roch relation (\ref{RR}) is applied both for $A_l$ and $D_l$ singularities.

  \section{One point function for $D_l$ type}
  
  The singularity theory of $D_l$ type is described as a two dimensional normal singularity by the equation of $x^{l-1} + x y^2 + z^2=0$ ($ l\ge 4$).
  The weight system is $(a,b,c;h) = (2,l-2,l-1; 2(l-1))$, where $h$ is called as Coxeter .  
  For $D_l$ case, we use the spin $p$ value for  Coxeter number  $h$, which is related to $l$ as $p= 2l -2$. 
  For $D_4$, we have spin curve of $p=6$. 
  
  The intersection numbers are extended from (\ref{Chern}) to the one includes the boundary. The tangent bundle is trivially  on the boundary
  of the moduli space. We need analogous term for the first Chern class on the boundary and introduce the correspondent quantity $\sigma$ for the boundary \cite{DijkgraafWitten}.
  \be
  <\tau_{n_1} \cdots \tau_{n_s} \sigma^m> = \int_{\bar {\mathcal M}} \psi^{n_1}\cdots \psi^{n_s}
  \ee
  where $m$ punctures on the boundary are added to (\ref{Chern}). This is generalization to open intersection numbers, and related to $D_l$ singularity
  as we will see. Instead of working of the  geometrical moduli space $\bar {\mathcal M}$, we study the equivalent partition function of a matrix model
  as same as $A_l$ case. The partition function of a matrix model is expressed by the $n$ point correlation function $u(s_1,s_2,...,s_n)$ with a logarithmic potential \cite{BrezinHikami5}.
  
  
  
  
  We have the following
  one point function $\hat u(s)$ in the integral form for $D_l$,
  \vskip 2mm
  The generating function $\hat u(s)$ of the intersection numbers $<\tau>$ for $D_l$  ($p= 2l-2$) is given by
  \be\label{ulog}
  \hat u(s) = \frac{1}{s}\int _0^\infty du e^{- \frac{c}{(p+1) }[ (u + \frac{s}{2})^{p+1}- (u - \frac{s}{2})^{p+1}]} \frac{1}{\sqrt{1 - \frac{s^2}{4 u^2}}}
  \ee
  where $c = \frac{N}{p-1}\sum \frac{1}{a_{\alpha}^{p+1}}$ as shown in \cite{BrezinHikami3}. $a_\alpha$ is eigenvalues of the external source.
  
  The last factor  is absent for $A_l$ case. It  is written as
  \ba\label{log1}
   \frac{1}{\sqrt{1 - \frac{s^2}{4 u^2}}} &=& \frac{1}{2}\biggl(\sqrt{\frac{u+ \frac{s}{2}}{u- \frac{s}{2}}} + \sqrt{\frac{u- \frac{s}{2}}{u + \frac{s}{2}}}\biggr)\nonumber\\
   &=& \frac{1}{2}\biggl( e^{\frac{1}{2}{\rm log} (u-\frac{s}{2})- \frac{1}{2}{\rm log} (u+ \frac{1}{2})} + e^{\frac{1}{2}{\rm log} (u+\frac{s}{2})- \frac{1}{2}{\rm log} (u- \frac{1}{2})}\biggr)
   \ea
Note that if we change $u \to - u$, above term is invariant. If we write the coefficient of the logarithm as $k$ instead of $\frac{1}{2}$, two terms are $k$ and $-k$ coefficient, and it leads to the polynomial of even power of $k$. This characterizes the $D_l$ type as we will discuss later.

    The small $s$ expansion of $u(s)$ with $u=t^{\frac{1}{p}}$, becomes
  \ba\label{integral}
  &&\hat u(s)= \frac{1}{N \pi s}\frac{1}{(cs)^{1/p}} \int_0^\infty dt t^{\frac{1}{p}-1} e^{-t} [ 1 - \frac{p(p-1)}{24} s^2 (cs)^{2/p}t^{1-\frac{2}{p}} \nonumber\\
  &&- \frac{p(p-1)(p-2)(p-3)}{5! \cdot 16} s^4 (cs)^{4/p} t^{1- \frac{4}{p}}
 +\frac{p^2(p-1)^2}{2\cdot (24)^2} s^4 (cs)^{4/p} t^{2-\frac{4}{p}}\nonumber\\
 &&- \frac{p^3(p-1)^3}{3! (24)^3} s^6 (cs)^{6/p} t^{3 - \frac{6}{p}}
 + \frac{p^2(p-1)^2 (p-2)(p-3)}{5!\cdot 14\cdot 16}s^6 (cs)^{6/p}t^{2-\frac{6}{p}}\nonumber\\
 &&-\frac{p^3(p-1)^3}{3!(24)^3}s^6 (cs)^{6/p} t^{3-\frac{6}{p}} \nonumber\\
 &&- \frac{p(p-1)(p-2)(p-3)(p-4)(p-5)}{7! 4^3}s^6 (cs)^{6/p} t^{1-\frac{6}{p}}+ \cdots]\nonumber\\
 &\times& [ 1+ \frac{1}{8} s^2 (cs)^{2/p}t^{-\frac{2}{p}} + \frac{3}{128}s^4 (cs)^{4/p}t^{-\frac{4}{p}}+ \frac{15}{3072} s^6 (cs)^{6/p}t^{-\frac{6}{p}} + \cdots]
  \ea
  The last factor is the expansion of (\ref{log1}). The integer power of $s$, denoted as $m$, shows the relation to genus $g$ as $2g-1 = m$.
  
   The small $s$ expansion of $\hat u(s)$, with the normalization factors $\frac{1}{p^{g-1}}$ and gamma factor $1/ \Gamma(1- \frac{1+j}{p})$, gives the intersection numbers
  of one point case, 
  \ba\label{tauD}
  <\tau_{1,0} >_{g=1} &=& \frac{p+2}{24}  \nonumber\\
  <\tau_{n,j } >_{g=2}  &=& \frac{(p+2)(2p+1)(p-6)}{5760 p} \frac{\Gamma(1- \frac{3}{p})}{\Gamma(1 - \frac{1+j}{p})} \nonumber\\
  <\tau_{n,j}> _{g=3} &=& \frac{(p+2)(2p+1)(8p^3 -77 p^2 + 196 p + 188)}{2903040 p^2} \frac{\Gamma(1-\frac{5}{p})}{\Gamma( 1- \frac{1+j}{p} )} \nonumber\\
  <\tau_{n,j}> _{g=4} &=& \frac{(p+2)(2p+1)(4p^2-27 p - 22)(18 p^3 - 133 p^2 + 308 p + 332)}{1393459200 p^3} \nonumber\\
  && \times \frac{\Gamma(1- \frac{7}{p})}{\Gamma(1 - \frac{1+j}{p})} \nonumber\\
  <\tau_{n,j}>_{g=5} &=& \frac{(p+2)(2p+1)(4p+3)(p-6)}{122624409600 p^4} (32p^5 -450 p^4 + 1741 p^3 \nonumber\\
  &&-1642 p^2
   -6788p-3288) \frac{\Gamma(1-\frac{9}{p})}{\Gamma(1-\frac{1+j}{p})} \nonumber\\
   <\tau_{n,j}>_{g=6} &=& \frac{(p+2)(2p+1)}{ 14! 6! 2^7 \cdot 3 p^5}(530688 p^9 -13260176 p^8 + 115768820 p^7\nonumber\\
   && -412604468 p^6 + 276695515 p^5 + 1715374838 p^4 -2129848328 p^3\nonumber\\
   &&-6843457424 p^2 -4961166736 p -1156803104)\frac{\Gamma(1- \frac{11}{p})}{\Gamma(1- \frac{1+j}{p})} \nonumber\\
   <\tau_{n,j}>_{g=7} &=& \frac{(p+2)(2p+1)}{16! 2^{10} 3^3 p^6} ( 2^{11}3^3 5 p^{11} -8586624 p^{10}+ 96830032 p^9\nonumber\\
   && - 488127956 p^8 + 889089716 p^7 + 1243914177 p^6 - 5937016268 p^5\nonumber\\
   &&-1741314004 p^4 + 21058826784 p^3+ 29690849392 p^2\nonumber\\
   && + 15502250816 p + 2905782080) \frac{\Gamma(1- \frac{13}{p})}{\Gamma(1- \frac{1+j}{p})} \nonumber\\
   <\tau_{n,j}>_{g=8} &=& \frac{(p+2)(2p+1)(p-6)(3+4 p)(5+ 6 p)}{18! 2^{15} 5 p^7} ( 462976 p^{10} \nonumber\\
   && - 15295120 p^9 + 179456596 p^8 - 953948892 p^7+ 2115904691 p^6\nonumber\\
   &&+ 586034636 p^5 - 9624755932 p^4 + 5128005728 p^3\nonumber\\
   &&+31236673872 p^2+ 25467952320 p + 6355800000 ) \frac{\Gamma(1- \frac{15}{p})}{\Gamma(1- \frac{1+j}{p})}  \nonumber\\
   <\tau_{n,j}>_{g=9} &=&  (2 + p) (1 + 2 p) (16502445084498176 + 122234441454621184 p \nonumber\\
   &&+ 
      362714955007461120 p^2 
      + 525312614038452992 p^3 \nonumber\\
      &&+ 
      326733211545349216 p^4 
      - 23742664329025152 p^5\nonumber\\
       &&- 
      100306262063206224 p^6 + 8323562725999632 p^7\nonumber\\
      && + 
      24146966038644009 p^8 - 6773140965548282 p^9\nonumber\\
      && - 
      2307841939577188 p^{10} + 1677474927489096 p^{11}\nonumber\\
      && - 
      402367617574016 p^{12} + 48227812538240 p^{13} \nonumber\\
      && - 
      2854331624448 p^{14} + 64684523520 p^{15})\nonumber\\
      &&\times \frac{1}{
  271211974879377138647040000 p^8} \frac{\Gamma(1-\frac{17}{p})}{\Gamma(1-\frac{1+j}{p})}
  \ea
  \ba\label{tauD2}
  <\tau_{n,j}>_{g=10} &=& (2 + p) (1 + 2 p) (-24830402547748278784 - 209477583844413564160 p \nonumber\\
     &&- 
     730557543123682249216 p^2 - 1318998112825968482560 p^3\nonumber\\
     && - 
     1208129128709188488640 p^4 - 333863005164255047776 p^5\nonumber\\
     && + 
     264649746344510118240 p^6 + 126068631741386072496 p^7 \nonumber\\
     &&- 
     78507951575573824290 p^8 - 22445884710909365885 p^9\nonumber\\
     && + 
     21279933896679839896 p^{10} - 1348924974460414280 p^{11} \nonumber\\
     &&- 
     2661096526358202656 p^{12} + 1087695793798917040 p^{13} \nonumber\\
     && - 
     200715490239800960 p^{14} + 19791837888816384 p^{15} \nonumber\\
     && -  993551356753920 p^{16} + 
     19465064349696 p^{17}) \nonumber\\
     &&\times \frac{1}{3579998068407778230140928000000  p^9}\frac{\Gamma(1- \frac{19}{p})}{\Gamma(1- \frac{1+j}{p})}\nonumber\\
     <\tau_{n,j}>_{g=11} &=& ((-6 + p)  (2 + p) (1 + 2 p) (3 + 4 p) (7 + 
      8 p)\nonumber\\
      &&\times (-467169353783096832 - 3341630372025717504 p\nonumber\\
      && - 
      9468101148218970624 p^2 - 12751291781631800064 p^3 \nonumber\\
      &&- 
      6632138067837797184 p^4 + 1788770829880213088 p^5 \nonumber\\
      &&+ 
      2276784479751975200 p^6 - 728458396311840240 p^7 \nonumber\\
      &&- 
      535146263253641670 p^8 + 300779490648921045 p^9 \nonumber\\
      &&+ 
      17161921882855788 p^{10} - 53942771498453544 p^{11} \nonumber\\
      &&+ 
      19196939796342336 p^{12} - 3360970337005104 p^{13} \nonumber\\
      &&+ 
      319429042188736 p^{14} - 15406349322752 p^{15} \nonumber\\
      &&+ 
      286370611200 p^{16})\nonumber\\
      &&\times \frac{1}{73191071620781243816214528000000 p^{10}}\frac{\Gamma(1-\frac{21}{p})}{\Gamma(1-\frac{1+j}{p})}
    \ea
   where $n$ and $j$ are determined by the Riemann-Roch selection rule of (\ref{RR}).
   
   When $p=6$, which corresponds to $D_4$ (Coxeter number $p$ is equal to $2l-2$ for $D_l$), the intersection numbers $<\tau>_{g}$ are vanishing for $g= 2+ 3 k$, $k\in Z$.
   This vanishing intersection numbers can be seen explicitly as a factor $(p-6)$ in above expression. It is interesting to observe that the factor $(p-6)$
   accompany the factor $(3+4 p)$ in the $g=5, 8, 11$   for both $A_{p-1}$ and $D_l$ cases. 
    \cite{BrezinHikami5}.
   The existing factors $(3+ 4 p)$ are expected to appear in higher genus both for $A_{p-1}$ and $D_l$, when $g= 5 + 3m, (m=0,1,2,...)$.
   The vanishing condition of $g = 2+ 3 k, k\in Z$ for $p=6$ will be discussed at later section in more details.
   
   We  find for $p=2$ that the simple expression for one point function  ($j=0$ for $p=2$) is obtained  from (\ref{tauD}),
  \be\label{oneD}
  <\tau_{3g-2}> = \frac{1}{g! 6^g}
  \ee
  Above formula for $p=2$ will be  proved  later in (\ref{proof}).

  There is a relation to the open intersection numbers. The logarithmic potential with a coefficient $k$ has been studied so called as Kontsevich-Penner model or Airy matrix model with
  a logarithmic potential \cite{BrezinHikami10,BrezinHikami4,Bertola}.
  Putting $p=2$ and $k=\frac{1}{2}$ in the expressions of (\ref{log}) \cite{BrezinHikami10,Bertola}, the results agree with (\ref{tauD}). 
  For instance, $<\tau_{1,0}>_{g=1} = \frac{p+2}{24}$ becomes $\frac{1}{6}$ for $p=2$, which agrees with $<\tau_{1}>= \frac{1+ 12 k^2}{24}= \frac{1}{6}$ with $k=\frac{1}{2}$ in \cite{BrezinHikami10,Bertola}. $<\tau_{n,j}>_{g=2}= \frac{1}{72}$ for $p=2$ agrees with $<\tau_{4}>= \frac{1 + 56 k^2 + 16 k^4}{1152} = \frac{1}{72}$ for $k=\frac{1}{2}$. 
  For non orientable surface, taking account of projective plane
  $(g=\frac{1}{2})$, Klein surface $(g=1)$, and crosscap $(g=\frac{3}{2})$, genus $g$ is considered as a fractional (double genus) \cite{BrezinHikami7}. Thus for the comparison with the expressions
 of $<\tau_{\frac{3g'-1}{2}}>$  in \cite{BrezinHikami4,Bertola}, we need a relation $g'= 2g-1$ i.e. the one point function of (\ref{tauD}) corresponds to $<\tau_{\frac{3g'-1}{2}}>$ in \cite{BrezinHikami4,Bertola}, and the result completely agrees for the case of integer $\frac{3g'-1}{2}$. The case $p=2$ in $D_l$ type means $D_2$, so $D_2$ case has a meaning of the Kontsevich-Penner model as 
 an open intersection theory, although  $l\ge 4$ case is discussed in the singularity theory. The terms of $<\tau_{\frac{3g'-1}{2}}>$, which appear with half-integer  $\frac{3g'-1}{2}$, have  expressions of odd
 $k$ polynomial, and they do not appear in $D_l$ type in (\ref{tauD}). 
 
     For the open intersection numbers, there appear intersection numbers such that $<\tau_{\frac{5}{2}}>= \frac{1}{12}(k + k^3)$. These are odd power of $k$. As we observed in (\ref{proof}),
     we have sum of $k$ and $-k$ for the logarithmic correction, therefore adding these two contributions, they are cancelled for such $<\tau_{\frac{5}{2}}>$ intersection numbers.
     This is a reason that we have no half integer indexed intersection numbers in $D_l$ type.
     
     As noted in \cite{BrezinHikami10}, such odd k intersection numbers with half integer indexed $\tau$ are evaluated by the contour integral, and therefore can be regarded as Ramond
     sectors.
 
 Thus the intersection numbers belong to the orientable surfaces, which do not include the case of half integer
 $\frac{3g'-1}{2}$.
 For general $p$, if we put $k= \frac{1}{2}$ in \cite{BrezinHikami10}, we find also the agreement with (\ref{tauD}).

   For $p=6$, which corresponds to $D_4$, the intersection numbers become $\tau_{D_4}= 1, \frac{1}{2}, 0, \frac{1}{40824}, \frac{13}{122472},
  0 $ for $g=0,1,2,3,4,5$, respectively. These values are obtained from (\ref{tauD}) as the coefficients of  Gamma function factors. They agree with \cite{Dubrovin}. 
  It is remarkable that $\tau_{D_4}$ is vanishing at $g=2, 5$, since a factor $(p-6)$ appears in (\ref{tauD}). This suggests $\tau_{D_4}$ ($p=6$ case) is vanishing periodically at $g=2+3 k$, $k \in Z$. 

    This periodic vanishing  reallation of the intersection numbers is due to  the selection rule for spin $p$.
  We have from Riemann-Roch formula  the relation between the spin $p$ and the genus $g$,
  \be\label{RR1}
  (p+1)(2g-1) = p n + j + 1
  \ee
  where $j = 0, 1, 2, ..., p-2$. For the singularity theory, $D_l$ has a relation
  \be\label{RR2}
  2g -1 = p n' + m_\alpha
  \ee
  where $m_\alpha$ is defined by the characteristic polynomial $\chi(t)$ for the weight system $(a,b,c;h)$ as
  \be
 \chi(t) = \frac{1}{t^h} \frac{(t^h - t^a)(t^h-t^b)(t^h- t^c)}{(t^a - 1)(t^b - 1) (t^c - 1)}= \sum_{\alpha=1}^l t^{m_\alpha}
 \ee
 with $l$ is Milnor number. $j$ is $m_\alpha - 1$, which is the spin components of spin $p$.
 
 For $D_4$, the weight system becomes $(2,2,3;6)$, and $m_\alpha= 1,3,5$ ($m_2=m_3=3$, double).
 The relation of (\ref{RR1}) is written by $n'= n+1$, $j= m_\alpha-1$ in (\ref{RR2}).
 For $p=6$ in $D_4$, $g=1$ corresponds to $m_\alpha= 1$ (j=0,n=1). $g=2$ corresponds to $m_\alpha= 3$ ($j=2,n=3$).
 $g=5$, $g=8$ correspond to $m_\alpha=3$. Thus for $g= 2+ 3 k$ ($k \in \mathbb{Z}$), the exponent becomes $m_\alpha= 3$, which is doubled.
 This correspondence is related to the vanishing relation of the intersection numbers of $D_4$ for $g = 2+ 3 k$.
   
  From (\ref{tauD} and (\ref{tauD2})), It is easily recognized ;  (i) large $p$ behavior is same as A-type (\ref{A}), (ii) the intersection numbers of  $p=-1/2$ are vanishing for $g>1$ for all order of $g$ (same as A-type), (iii) the intersection numbers of $p=-2$ for D-type are vanishing for all genus $g>0$. These remarkable properties will be proved in the following sections.
  
    Since we have derived exact one point function of $D_l$ type, it may be interesting to consider the negative values of $p$ and half-integer $p$ as discussed in  $A_l$ type \cite{BrezinHikami5,BrezinHikami1, BrezinHikami2}. In the next section, we examine the negative integer values of $p$ and half-integer $p$. Interesting applications of such non-positive integer
    cases were discussed in \cite{BrezinHikami2} for $A_l$ type.
  \vskip 2mm
 
 \vskip 2mm
  \section{One point function of $D_l$ type for the non-positive integer cases}
\vskip 2mm

We have discussed the non-positive integer spin $p$ for one marked point of the $A_{p-1}$ type in the previous articles \cite{BrezinHikami1,BrezinHikami2}.
Here we extend these results of $A$ type to $D$ type for one marked point. 

 For $D_l$ singularity,
 the relation of $p= 2l-2$ gives the constraint that spin $p$ should be even integer. 
Since the intersection numbers are expressed by the polynomial of $p$ as (\ref{tauD}), the analytic continuation of $p$ to the general values including the non-integer case is possible. Our formulation of a matrix model allows the non-positive integer value of $p$. 

In this section, we will find the remarkable coincidence of the intersection number  of $D_l$ type with that of $A_{p-1}$ model with a logarithmic term, so called generalized Kontsevich-Penner model.
\vskip 2mm
{\bf{$\bullet$ Change of variable}}
  \vskip 3mm
  
  As discussed in \cite{BrezinHikami1,Fan}, although there is no Ramond contribution for $A_{p-1}$ case in the positive integer $p$,   there appear Ramond punctures in $D_l$ type ($p= 2l-2$) \cite{BrezinHikami2,Fan}. The Ramond contribution may be obtained
  by the residue of $y=0$ in the following integral representation by a change of variable  from $u$ to $y$,  following the discussion of \cite{BrezinHikami1,BrezinHikami2} as
  \be\label{y}
  u = \frac{i}{2}( y^2 - \frac{1}{y^2})
  \ee
  The factor of $D_l$ logarithmic potential becomes after the change of variable of (\ref{y}),
 \be\label{yy}
  \frac{u}{\sqrt{u^2-1}}= \frac{y - \frac{1}{y^3}}{y + \frac{1}{y^3}}
  \ee
  The measure for $A_{p-1}$ type  is now changed  simply to $ i  (y+ 1/y^3) dy$ due to (\ref{yy}), which reads that the measure is $y \pm \frac{1}{y^3}$, $(+)$ sign for A-type and $(-)$ sign
  for D-type.
  \vskip 2mm
  {\bf$\bullet$ Equivalence to generalized Kontsevich-Penner model of open intersection numbers}
  \vskip 2mm
  Extension of Airy with logarithmic potential (Kontsevich-Penner model)  to general $p$ spin case has been investigated
 with the logarithmic potential with coefficient $k$ in  \cite{BrezinHikami4,BrezinHikami10}.
 We will show that  the intersection numbers of generalized Kontsevih-Penner model with $k=\pm \frac{1}{2}$ are identical to that of $D_l$ ($p=2l-2$) intersection numbers.

For genus $g=1$, the intersection number is
\be\label{generalk}
<\tau>_{g=1} = \frac{p-1 +12 k^2}{24}
\ee
With $k=\pm \frac{1}{2}$, it becomes 
\be
<\tau> = \frac{p+2}{24}
\ee
which agrees with $\tau_{D_l}$in (\ref{tauD}) . For genus $g=2$, the generalized Kontsevich-Penner model gives (Eq.(5.28) of \cite{BrezinHikami4}),
\be
<\tau>_{g=2} = \frac{1}{p(12)^2} [\frac{(p-1)(p-3)(2p+1)}{40} - (3p+1) k^2 - 2k^4]
\ee
 By putting $k=\pm \frac{1}{2}$, we find exactly $<\tau>_{g=2}  = \frac{1}{5760 p} (p+2)(2p+1)(p-6)$ for D-type in (\ref{tauD}).
 
 Thus we find that the one point intersection number of $D_l $ type ($p=2l-2$) is same as the intersection numbers of generalized Kontsevich-Penner matrix model with $k=\frac{1}{2}$.
 One can check more higher $g$ case are consistent with this identification. When $p=2$, the generalized open intersection numbers are  evaluated in higher orders \cite{Bertola}, in which a parameter $N$ is same as our $k$.  The odd power terms of $k$ are cancelled by adding
  $\pm k$ contributions, and the results agree with $D_l$ type intersection numbers with $k=\frac{1}{2}$. Note that the open intersection number is described by the logarithmic potential with $k=1$ \cite{BrezinHikami10,BrezinHikami4,BrezinHikami3, Pandharipande, Buryak, Alexandrov1}.

\vskip 2mm
  {\bf{$\bullet$ $p=1$ case}}
  \vskip 2mm
  We now discuss and prove the remarkable features for specific values of $p$.
  The first example is  $p=1$ case of $A_l$. The intersection number $<\tau>$ becomes vanishing in all order of $g$ due to a factor$(p-1)$ in (\ref{A}).
  \ba
  u(s) 
  &=& \frac{1}{s}\int du e^{-\frac{c}{2}((u+\frac{s}{2})^2-(u-\frac{s}{2})^2)}\nonumber\\
  &=& \frac{i}{s}\oint \frac{dy}{2 i \pi} (y +\frac{1}{y^3}) e^{-\frac{ics}{2}  (y^2- \frac{1}{y^2})}\nonumber\\
  &=& \frac{i}{s} \oint \frac{dy}{2 i \pi} (\frac{1}{- i c s}) \frac{d}{dy} e^{-\frac{ics}{2}  (y^2- \frac{1}{y^2})}
  \ea
  which becomes vanishing due to the total derivative. This is consistent with a factor $(p-1)$ in all order in (\ref{A}).
  
  For $D_l$ case, the measure is a factor $(y - \frac{1}{y^3})$ instead of $(y + \frac{1}{y^3})$, 
  \ba
  u(s) &=& \frac{i}{s}\oint \frac{dy}{2 i \pi} [2 y - (y + \frac{1}{y^3})] e^{-\frac{ic s}{2}  (y^2- \frac{1}{y^2})}\nonumber\\
  &=& - \sum_{m=0}^\infty \frac{1}{2^{2m} m! (m+1)!} s^{2m} c^{2m+1}
  \ea
  This is consistent with  the expression of (\ref{tauD}) for $p=1$. The genus $g$ is equal to $m$.
  
\vskip 2mm
  {\bf{$\bullet$ $ p=2$ case}}
  \vskip 2mm
  
   For $p=2$, using the representation of $y$, one point function $u(s)$ is expressed as
  \be
  u(s) = \frac{i}{2} e^{-\frac{5 c}{24}s^3 }  \oint \frac{dy}{2i\pi} (y \pm \frac{1}{y^3}) e^{\frac{cs^3}{16} (y^4 + \frac{1}{y^4})}
  \ee
  where $\pm$ means $A$ type and $D$ type, respectively. With the change of variable $y = t^{\frac{1}{4}}$, this is written by the modified Bessel function $I_\nu(z)$,
  \ba\label{proof}
  u(s) &=& \frac{i}{8}  e^{-\frac{5 c}{24}s^3 } \oint \frac{dt}{2 i \pi} (t^{-\frac{1}{2}}\pm t^{-\frac{3}{2}})e^{\frac{cs^3}{16}(t + \frac{1}{t})}\nonumber\\
  &=&\frac{i}{8}  e^{-\frac{ 5 c}{24}s^3 } [ I_{-\frac{1}{2}}( \frac{cs^3}{8}) \pm I_{\frac{1}{2}}( \frac{cs^3}{8}) ]\nonumber\\
  &=& \frac{i}{2}  e^{-\frac{5 c}{24}s^3 }  \sqrt{\frac{1}{\pi ( cs^3)}} e^{\pm \frac{cs^3}{8}} 
  \ea
  By taking $c= -\frac{1}{2}$, we find  the close form of for $A$ type (\ref{Kontsevich}); $<\tau_{3g-2}>= \frac{1}{g! (24)^g}$, and the result of D-type (\ref{oneD}); $<\tau_{3g-2}> =\frac{1}{g! 6^g}$. This gives a proof of (\ref{oneD}).

  \vskip 2mm
  {\bf{$\bullet$ $p = - 1$}}
  \vskip 2mm
  When $p=-1$, one point function $u(s)$ provides Euler characteristic $\chi(\bar M_{g,1})$ for $A_l$ type singularity \cite{BrezinHikami3} as
  \ba\label{chi}
  u(s) &=& \frac{1}{N} \int \frac{du}{2i\pi} \biggl(\frac{u-\frac{1}{2}}{u+\frac{1}{2}}\biggr)^N \nonumber\\
  &=& -\frac{1}{N}\int_0^\infty \frac{dz}{2\pi} \frac{e^{-z}}{(1-e^{-z})^2} e^{-Nz}\nonumber\\
  &=& \int_0^\infty \frac{dz}{2\pi}(\sum B_{2n} \frac{z^{n-1}}{n!}) e^{-Nz}
  \ea
  where a change of variable $(u-\frac{1}{2})/(u+\frac{1}{2}) = e^{-z}$ is used and a factor $1/N^2$ represent  the genus $g$ expansion. This $u(s)$ gives Euler characteristic $\chi(\bar M_{g,1}) = \zeta(1-2g) = (- 1)^{g}\frac{1}{2g}B_{2g}$, where $B_n$ is a Bernoulli number ($B_2= \frac{1}{6}, B_4= - \frac{1}{30}, B_6= \frac{1}{42}, ...$).

  For $D_l$ type, if we put $p=-1$ in (\ref{tauD}) with $j=0$ (gamma function of the denominator becomes one for $p=-1, j=0$, and numerator gamma function gives $(2g-1)!$),
  , we find that $<\tau>_g $ as $\frac{1}{24}, -\frac{7}{960}, \frac{31}{8064}, ...$ for $g=1,2,3,...$. These numbers are equal to Euler characteristics $\chi$
  
  \be\label{chiD}
  \chi = (1 - 2^{1-2g})\frac{B_{2g}}{2g}. 
  \ee
  
  In the limit $p\to -1$, (\ref{ulog}) becomes
  \ba
  \hat u (s) &=& \frac{1}{2}\int du e^{-c log ( \frac{u+1}{u-1})} \frac{u}{\sqrt{u^2 - 1}}\nonumber\\
  &=& \frac{1}{2}\int du (\frac{u-1}{u+1})^c \frac{u}{\sqrt{u^2-1}}
  \ea
  By the change of variable $(u-1)/(u+1) = e^{-z}$, $du = -2 e^{-z}/(1- e^{-z})^2 dz= (-2) (e^{z/2}- e^{-z/2})^{-2} dz$, $u/\sqrt{u^2-1} = \frac{1}{2}(e^{z/2}+ e^{-z/2})$, it becomes
  after partial integration, 
  \ba
  \hat u(s) = \int dz \frac{1}{e^{z/2}- e^{-z/2}} e^{-Nz}
  \ea
  where $c$ is replaced by $N$ to make clear of genus dependence, as $\frac{1}{N^{2g}}$ series. Above integral reduces to (\ref{chiD}), with $1/24$, 7/960,... for $g=1,2,...$.
  This $\chi$ is same as virtual Euler characteristics, obtained for $o(2N)$ matrix model  as a non-orientable surface \cite{BrezinHikami5}.
  For $p=-1$, there is no $\psi$ class, and only Euler class (Witten class) exists. Since $D_l$ type is related to $o(2N)$ Lie algebra, it is reasonable to obtain the result of (\ref{chiD}),
  which is same as a virtual Euler characteristics of real algebraic curves \cite{Goulden}. Indeed, we have obtained this virtual Euler characteristic of real algebraic curves for
  $so(2n+1),so(2n), sp(n)$ cases as antisymmetric matrix models \cite{BrezinHikami7,BrezinHikami5}.

  \vskip 3mm
  {\bf $\bullet$  $p$ = $-\frac{1}{2}$ }
  \vskip 2mm
  It is remarkable that the intersection number has a factor of $(2p+1)$ for $D_l$ type, which is same as  $A_l$ case  (\ref{A}). The reason of this factor $(2 p+1)$ exists for all genus ($g > 1$) is explained   by the change of variable of (\ref{y}).
   The exponential factor in (\ref{ulog}) becomes simply as $e^{- c\frac{s^{1/2}}{y}}$ for $p=-1/2$.   Therefore we have for $D_l$ case with $p=-\frac{1}{2}$ from (\ref{yy}),
  \be
  \hat u(s) = \oint_c \frac{dy}{2i \pi}  (y - \frac{1}{y^3})  e^{- c \frac{s^{1/2}}{y}}
  \ee
   where the integral is evaluated by a contour around $y=0$. This gives non-vanishing term of order $s$, which means $g=1$, but the remaining terms of all higher genus ($g > 1$)  should vanish due to Cauchy theorem.
   
   \vskip 3mm
   {\bf{$\bullet$ $p = - 2$}}
   \vskip 2mm
   In  $A_{p-1}$ matrix model, the case $p=-2$  becomes equivalent to the unitary matrix model  (BGW) model \cite{BG,GW}, and the intersection numbers agree with BGW matrix model as shown in \cite{BrezinHikami8}.
   
   For $D_l$ ($p = 2l -2$) case, there is  a factor $(p+2)$ for all genus case as  in (\ref{tauD}).     This may be explained as follows \cite{BrezinHikami8}.
   We have in the  $p= - 2$ case $l=(p+2)/2=0$ for $D_l$, namely $D_0$ case,
   \ba\label{k}
   \hat u(s) &=& \frac{1}{s} \oint \frac{du}{2i \pi} e^{-\frac{1}{u+s} + \frac{1}{u}} \biggl( \frac{u+ s}{u}\biggr)^k \nonumber\\
   &=&   - \frac{1}{4 \pi \sqrt{s}}\int dx \frac{1}{x^2} e^{-\frac{4 x^2}{1 + s x^2}}(\frac{1 - i x \sqrt{s}}{1+ i x \sqrt{s}})^k \nonumber\\
   &=& -\frac{1}{2\sqrt{\pi s}}(- s \frac{1}{8}(4k^2-1)+ s^2 \frac{1}{3!2^7} (4k^2-1)(4k^2-9) \nonumber\\
   && - s^3 \frac{1}{5! 2^9}(4k^2-1)(4k^2-9)(4k^2- 25)  + \cdots)
   \ea
   The second line is derived by the change of variable $u = \frac{1}{2} (-1 + \frac{i}{x \sqrt{s}})$. 
   Since the coefficient of the logarithmic factor of $D_l$ is  $k=\frac{1}{2}$, we find the vanishing results of $\hat u (s)$ from above equation. This gives a proof  for a factor $p+2$ for the intersection numbers in all orders of D-type (\ref{tauD}).
    
    BGW model is considered for the unitary matrix, and its extension to O(N)  group was studied in \cite{HikamiMaskawa}. There appears a phase transition between
    weak coupling (small $s$) and a strong coupling regions (large $s$).
    
   \vskip 2mm
   {\bf{$ \bullet$ Large $s$ expansion for $p= -2$}}
   \vskip 2mm
  The large $s$ expansion was investigated in \cite{BrezinHikami8} based on (\ref{k}) for $p=-2$ of $A_l$ type. Due to two terms of (\ref{log}), the extension of the large $s$ expansion for $p=-2$ case of $D_l$
  type is straightforward.
  We obtain by the shift $u\to (u-1)/2$ following \cite{BrezinHikami8},
  \ba
  \hat u(s) &=& \frac{1}{2}   \oint \frac{du}{2i \pi} e^{\frac{4}{2(u^2-1)}} [ (\frac{u+1}{u-1})^k + (\frac{u+1}{u-1})^{-k} ]\nonumber\\
 &=& \frac{1}{2}\sum_1^\infty \frac{4^m}{m! s^m} \oint \frac{du}{2 i \pi} \frac{1}{(u^2 - 1)^m} [ (\frac{u+1}{u-1})^k + (\frac{u+1}{u-1})^{-k} ]
  \ea
with $k= \frac{1}{2}$.
Noting that
\ba
&&\oint \frac{du}{2 i \pi} \frac{1}{(u^2-1)^m}(\frac{u+1}{u-1})^k = - \frac{2}{\pi}{\rm sin}\pi k \int_1^\infty dx \frac{(x+1)^{k-m}}{(x-1)^{k+m}}\nonumber\\
&&= - \frac{2}{\pi}({\rm sin}\pi k ) 2^{1-2m}(2m-2)!\frac{\Gamma(-k-m+1)}{\Gamma(-k+m)}
\ea
  we obtain $\hat u(s)$ in the large $s$ expansion. In the unitary matrix model, we put $k= -N$ and the result agree with the strong coupling expansion (character expansion) \cite{BrezinHikami8}.
  \vskip 3mm
  
  {\bf{$\bullet$ Half-integer $p=\frac{1}{2}$ and   $\frac{3}{2}$}}
  \vskip 2mm
  For $D_l$ type, with the change of variables $ u \to y$ in (\ref{y}), one point function $\hat u(s)$ is
  \be\label{1/2}
  \hat u(s) = \frac{1}{s} \oint \frac{dy}{2i \pi} (y-  \frac{1}{y^3}) g(y)
  \ee
    Note that the difference between $A_l$ and $D_l$ is a factor of $(y \pm \frac{1}{y^3})$, where $(+)$ is for $A_l$ and $(-)$ is for $D_l$. 
    The factor $g(y)$ is same as $A_l$ case. We have for $A_l$,
    \be
    u(s) = \frac{1}{s}\int \frac{du}{2 i \pi} e^{-\frac{1}{(p+1) } [ (u + \frac{s}{2})^{p+1} - (u - \frac{s}{2})^{p+1}]}
    \ee
    By the change of variable of $u = \frac{i}{2}(y^2-y^{-2})$, it becomes after the replace $u \to s u/2$, we have for $A_l$ case $(+)$ and $D_l$ case $(-)$
    \be\label{pm}
    u(s) = \frac{i}{2}\int \frac{dy}{2i \pi} (y \pm \frac{1}{y^3}) e^{- \frac{s^{p+1}}{(p+1)} (\frac{i}{4})^{p+1} \frac{1}{y^{2(p+1)} }[ (y^2-i)^{2p+2}-(y^2+i)^{2p+2}]}
    \ee
    where the exponential factor $g(y)$  is expressed as
  \be
  g(y) = e^{c' s^{3/2} (3y - \frac{1}{y^3})}  \hskip 2mm (p=\frac{1}{2},c'= -\frac{1}{6}i^{1/2})
  \ee
  The half-integer $p$ case  is related to Ramond sector \cite{BrezinHikami1,BrezinHikami2}.
  We assume Riemann-Roch relation for $p=\frac{1}{2}$ is valid as
  \be\label{RRb}
  3g-3 + 1 = n + (g-1)(1-\frac{2}{p}) + \frac{j}{p}
  \ee
  with $p= \frac{1}{2}$ and $j=-1$. $j=-1$ means Ramond puncture \cite{BrezinHikami2}. This selection rule indeed valid for the small $s$ expansion of $u(s)$.
  We have from (\ref{RRb})
  \be\label{n}
  n= 6g -3
  \ee
  which means that $u(s)$ is a power series of $s^{6g-3}$ and the intersection number is given by the coefficient, $<\tau_{n,j}>_g = <\tau_{6g-3,-1}>_g$.
  
  The one point function $u(s)$ for $p=\frac{1}{2}$ for $A_l$  is evaluated with the integral by part,
  \ba\label{onepoint1/2}
  u(s) &=& \frac{i}{2}\oint \frac{dy}{2 i \pi} (y + \frac{1}{y^3}) e^{c' s^{3/2} (3 y - \frac{1}{y^3})} =
  \frac{i}{6c 's^{3/2}} \oint \frac{dy}{2i\pi} y \frac{d}{dy} e^{c' s^{3/2} (3y- \frac{1}{y^3})}\nonumber\\
  &=& -\frac{i}{6c 's^{3/2}} \oint \frac{dy}{2i\pi}  e^{c' s^{3/2} (3y- \frac{1}{y^3})}\nonumber\\
  &=& 2 \sum_{g=1}^\infty \frac{(-1)^g}{g! (3g-1)!} (\frac{1}{48})^g s^{6g-3}
  \ea
  with $c' = - i^{1/2}/6$. The summation is over genus $g$ due to (\ref{n}).
 
   This generating function of the intersection numbers shows the precise agreement with (\ref{A}). Note that $\Gamma(1- \frac{2g-1}{p}) $ in (\ref{A}) is changed to $\Gamma(-1)$ by multiplication factor, and this $\Gamma(-1)$ is interpreted  as a normalization factor for the case of  $p=\frac{1}{2}$. Then we find the precise agreement for $p=\frac{1}{2}$  case between (\ref{onepoint1/2}) and (\ref{A}). 
  The case of $p=\frac{1}{2}$ is interpreted as
  a manifestation of Ramond puncture since spin component is $j= -1$ (The denominator $\Gamma(1- \frac{1+j}{p})$ in (\ref{A}) becomes one). This fascinating result of $p=\frac{1}{2}$ 
  will be further discussed in the next section related to the denominator of Bernoulli numbers. The intersection numbers are rational numbers and the denominator is common denominator of Bernoulli numbers. In $p=\frac{1}{2}$ case, there appears cancellation of this numbers of (\ref{denominator1}).

  For $D_l$ type ($p= 1/2, l= 5/4$),   it becomes from (\ref{yy})
  \ba\label{Dfactor}
  \hat u(s) &=& \frac{i}{2} \oint \frac{dy}{2 i \pi} (y- \frac{1}{y^3}) e^{c s^{3/2} (3 y - \frac{1}{y^3})}\nonumber\\
  &=& \frac{i}{2s} \oint \frac{dy}{2 i \pi} (y- \frac{s^2}{y^3}) e^{c s ( 3 y - \frac{s^2}{y^3})} 
  \ea
  This integral is written as the derivative of the exponent by $s$. Using the same integral as (\ref{onepoint1/2}), we obtain
  \ba
  \hat u(s) &=& \frac{i}{6cs}\frac{d}{ds}\biggl( \oint \frac{dy}{2 i \pi} e^{3 cs y - \frac{c s^3}{y^3}}\biggr)\nonumber\\
  &=& 2 \sum_{g=1}^\infty 
  \frac{(-1)^g (6g-1)}{g!(3g-1)!} (\frac{1}{48})^g s^{6g-3}
    \ea
    where $c= -\frac{1}{6} i^{\frac{1}{2}}$.
  This result is consistent for $p=\frac{1}{2}$ in (\ref{tauD}) as $<\tau>_{g=1} = \frac{p+2}{24}
  = \frac{5}{48}$.  The expression of $p=\frac{1}{2}$  for $D_l$ ($l= \frac{1}{2}(p+2)= \frac{5}{2}$) is obtained
  solely from the residue calculation, which means that this case is   Ramond puncture  with $j=-1$ \cite{BrezinHikami2}.

  For $p=\frac{3}{2}$ of $A_l$ type, it corresponds to $\beta\gamma$ system as discussed in \cite{BrezinHikami1}. There are two different punctures belongs to  Neveu-Schwarz and Ramond sectors. From (\ref{pm}), $D_l$ case is expressed as
  \be
   g(y) = e^{cs^{3/2} (5 y^3-\frac{10}{y}+ \frac{1}{y^5})} \hskip 2mm (p=\frac{3}{2}, c= \frac{4i}{5}(\frac{i}{4})^{5/2})
   \ee
  \be
  u(s) = \oint \frac{dy}{2i\pi} (y - \frac{1}{y^3}) e^{c s^{\frac{3}{2}}[5 y^3- \frac{10}{y} + \frac{1}{y^5}]}
  \ee
  The small $s$ expansion gives $<\tau_{n,j}>_g s^{n + \frac{2}{3}(j+1)}$, where $j=-1,- \frac{1}{2},0$. The spin $j=-1$ corresponds to
  Ramond sector, and Neveu-Schwarz sector is $j=-\frac{1}{2}, 0$. The Ramond sector (R) is evaluated by the contour integral for $A_l$ of $p=\frac{3}{2}$ in \cite{BrezinHikami1},
  \be
  u_R(s) = \frac{5}{2^5} c^2 s^5 - \frac{7}{3}\cdot \frac{5^4}{2^{16}} c^6 s^{15} + \frac{79\cdot 11}{10!}\cdot \frac{3^2 5^7}{2^{24}} c^{10} s^{25} +\cdots
  \ee
  where $s^{(2g-1)(1+ \frac{1}{p})} = s^{n+ \frac{1}{p}(j+1)}$. If we  take $p= \frac{3}{2}$ and $j=-1$, then we have $s^{\frac{10}{3}g - \frac{5}{3}}$.  The first term of above equation is for $g=2$, and the second term is for $g=5$.
    For $D_l$, $p=\frac{3}{2}$, the contribution of Ramond  sector becomes
    \be
    u_R (s) = \frac{105}{2^5} c^2 s^5 -\frac{19\cdot 17\cdot 5^4}{2^{19}} c^6 s^{15} + \frac{29\cdot 1663\cdot 5^5}{3^2 2^{32}} c^{10} s^{25} + \cdots
    \ee
    \vskip 3mm

        \section{Large $p$,  large $g$ limits and  integrality}
  \vskip 3mm
          In a recent paper \cite{Dubrovin}, the asymptotic behaviors of the intersection numbers for $A_{p-1}$, $D_l$ and $E_6$ types for the large $g$ are discussed based on the ordinary differential equations (ODE). In this section, we consider the large $g$ limit of $A_{p-1}$ and $D_l$ types based upon the integral representation of the intersection numbers $u(s)$, which may be simpler than the analysis of  ODE.
          
          The exponential parts of $u(s)$ of $A_{p-1}$ and $D_l$ are same. We write the exponent as a function $f(u)$, which becomes
          \be
          f(u) = c ((u+ \frac{s}{2})^{p+1} -(u-\frac{s}{2})^{p+1})
          \ee
          This function $f$ is a polynomial of $u$, hence this is an algebraic relation \cite{Dubrovin}. The one point function $u(s)$ is a series of $s^{(1+\frac{1}{p}) (2g-1)}$. Thus, large $g$ limit is equivalent to
          the large $s$ limit, and this limit is obtained by a saddle point method for the exponent $f(u)$. By the scaling $u \to \frac{s}{2} u$, $f(u)$ becomes
          \be
          f(\frac{su}{2}) = c (\frac{s}{2})^{p+1} ((u+ 1)^{p+1}-(u-1)^{p+1})
          \ee
          We denote this exponent as $g(u)$. By the saddle point method, the first derivative of $g$ is vanishing,
          \be
          \frac{d g}{du} = c (p +1) (\frac{s}{2})^{p+1} ((u+ 1)^p-(u-1)^p) = 0
          \ee
          which reads to 
          \be\label{circle}
          (\frac{u+1}{u-1})^p = 1
          \ee
          The solution of this equation is $\frac{u+1}{u-1}= e^{2\pi i/p}$, i.e. $u= (e^{\pi i/p}+ e^{-\pi i/p})(e^{\pi i/p} - e^{-\pi i/p})$.
  Thus, $f(us/2) = - c s^{p+1} ( 2 i {\rm sin} \frac{\pi}{p})^{-p}$. The intersection number $<\tau>_g$ is written by a contour integral of $t$ ($t= s^{2+ 2/p}$) in the large $g$ limit apart from Gamma function of the definition in (\ref{cs}), 
  \be
  <\tau>_g =  \oint \frac{dt}{2\pi i} \frac{1}{t^{g+1}} t^{-1/2} e^{- c t^{p/2}(2 i {\rm sin} (\pi/p))^{-p}}
  \ee
  where $t^{-1/2}$ factor comes from $1/s^{1+ \frac{1}{p}}$ in the front of the integral of $u(s)$, which corresponds to genus zero contribution.
   We expand the exponential term as $\sum \frac{1}{m!} (f(su/2))^m$, with
    $\frac{pm}{2} -\frac{1}{2}= g$.
   We have an asymptotic behavior in the large $g$ limit as
   \be\label{asymptotic}
   <\tau>_g \sim \frac{2 (p+1)^{\frac{1}{p}}{\rm sin}\frac{\pi}{p}}{(\frac{2g-1}{p})!} (\frac{1}{ 4 p (p+1)^{\frac{2}{p}} ({\rm sin} \frac{\pi}{p})^2})^g
   \ee
   where $c= \frac{1}{p+1}$ and the normalization factor $\frac{1}{p^{g-1}}$ is included. Thus the term of power $(-g)$ becomes $[(p+1)^{\frac{2}{p}} 4 p ({\rm sin}\frac{\pi}{p})^2 ]^{-g}$,
   which agrees with known $\frac{1}{(24)^g}$  for $p=2$ case.
   
   For $D_l$ ($p= 2l-2$) type, the exponential term is same as $A_{p-1}$. For the large $g$ behavior, the power $g$ part $a^g$ becomes same.
   \vskip 3mm
   
   Note that for $p= \frac{1}{2}$, there is no finite saddle point solution of (\ref{circle}),  except $u=\infty$. The asymptotic term of (\ref{asymptotic}) diverges for $p=\frac{1}{2k}, ( k\in Z$).  For $p=\frac{1}{2}$, we have used the change of variable from $u$ to $y$, and have obtained the explicit form of (\ref{onepoint1/2}) for the large $g$ limit.
   It is remarkable that the polynomial of $p$ of order $p^g$ in the intersection numbers of genus $g$ has all real roots. Namely the zeros of the polynomial are on the real
   axis.
   
    The values of this polynomials at $p=\frac{1}{2}$ are
   $-\frac{1}{2}, 5, (- 3^2\cdot 5\cdot 7), (3\cdot 4\cdot 5\cdot 7\cdot 13)$ for $g=1,2,3,4$, respectively. These numerators cancel with the denominator of Bernoulli numbers in (\ref{denominator1}), and the intersection numbers of $p=\frac{1}{2}$ is expressed simply as (\ref{onepoint1/2}). This fascinating feature of $p=\frac{1}{2}$ case shows that there is a characteristic topological meaning of the curves of Riemann surface for the fermionic spins.

        The intersection numbers $<\tau>_g$ are rational numbers since they involve the inverse of automorphism from  the orbifold. As remarked by Zagier \cite{Zagier}, there is  a property of the integrality by multiply certain factors of two Pochhammer symbols to the intersection numbers $<\tau>_g$. The denominators of the expression of $<\tau>_g$ are same
        for $A_{p-1}$ type and $D_l$ type as (\ref{A}) and (\ref{tauD}). Thus the integrality can be obtained by the application of the same Pochhammer symbols for both cases.
         
         As an example of integrality, with the multiplication of Pochhammer symbol $(x)_n = x (x+1) \cdots (x+n-1)$ to $<\tau>_{g = 5n}$, the following quantities  become  \cite{Dubrovin,Zagier,Bertola}.
         
         \ba\label{an}
         a_n &=&  (2^{10} 3^5 5^2)^n (\frac{3}{5})_n (\frac{4}{5})_n  <\tau>_{g=5 n}\nonumber\\
         b_n &=& (2^{12}3^5 5^4)^n (\frac{2}{5})_n (-\frac{1}{10})_n <\tau>_{g=5n} 
         \ea
        which are integers (apart from the normalization of $\frac{1}{p^{g-1}}$ ), where $<\tau>_{g=5}= 161/777600000$ for $A_4$ ($p=5$) in (\ref{A}).
        The generating function $\sum b_n t^n$ is algebraic, while $ a_n $ grows exponentially \cite{Zagier}.
        
        As evaluated in \cite{BrezinHikami13,BrezinHikami5}, the higher spin $p$ limit ($ p \to \infty$) of each intersection numbers of a fixed genus $g$ shows interesting features, which are expressed by Bernoulli numbers $B_{2g}$. From (\ref{A}) and (\ref{tauD}), it is easily noticed that the values of the coefficients of highest order of $p$ become same for $A_{p-1}$ and $D_l$ case for a fixed genus $g$.
        \be\label{largep}
        {\lim_{p\to \infty}} <\tau>_{g} |_{A_{p-1}}= {\lim_{p\to \infty}} <\tau>_{g} |_{D_{l}} = \frac{(-1)^{g+1}B_{2g}}{(2g)!(2g)} p^g + O(p^{g-1})
        \ee
       The denominators of the expression of $<\tau>_g$ are same, and then integrality should be same for $A_{p-1}$ and $D_l$. 
         The result of (\ref{largep}), which reduces to Bernoulli number $B_n$, has been obtained in \cite{BrezinHikami13}, and the relation to the partition function of black hole is discussed.  With $\sigma= \frac{s}{p}$, $u^{p+1}= x^2$, $u(s)$ is written in the large $p$,
         \ba\label{largeplimit}
         u(s) &=& \frac{2}{\sigma} \int \frac{dx}{2 i \pi} x^{-1+ \frac{2}{p}}e^{- \frac{2 c}{p+1} x^2 {\rm sh} \frac{\sigma}{2}}\nonumber\\
         &=& \frac{2}{\sigma} \Gamma(\frac{2}{p}) (\frac{2c}{p+1})^{-\frac{1}{2}}
         (\frac{\sigma}{2})^{-\frac{1}{p}} {\rm exp}(- \frac{1}{p}{\rm log}  \frac{{\rm sh} \frac{\sigma}{2}  }{ \frac{\sigma}{2} } )
         \ea
         Noting that ${\rm log}  \frac{{\rm sh} \frac{\sigma}{2}  }{ \frac{\sigma}{2} } = \sum(-1)^{n-1}\frac{B_{2n}\sigma^{2n}}{(2n)! 2n}$, the Bernoulli number $B_{2n}$ is obtained
         for the intersection number in the large $p$ limit.
         
         The notation of Bernoulli
         numbers are $B_2= \frac{1}{6}, B_4= -\frac{1}{30}, B_6= \frac{1}{42}, B_8= - \frac{1}{30}, B_{10}= \frac{5}{66}, B_{12}= - \frac{691}{2730}, B_{14}= \frac{7}{6},
         B_{16}= - \frac{3617}{510}$. The denominator of $<\tau>_g$ is common in the denominator of $B_{2g}/((2g)! (2g))$ for arbitrary $p$, since the intersection number is described
         by the multiplication of polynomial of $p$ to $B_{2g}/((2g)! (2g))$ for $A_l$ and $D_l$ cases.
         
         The denominator of $B_{2g}/(2g)$ is given by \cite{Mazur,Milnor}
         \be\label{denominator1}
         {\rm {denominator}} \bigl(\frac{B_{2g}}{2g}\bigr) = \prod_{p>2, (p-1)|2g} p^{1+ v_p(2g)}
         \ee
         The condition of the prime $p$ is $(p-1)|(2g)$, which means the $(p-1)$ divides $(2g)$, and $v_p(N)$ is the largest exponent $e$ such that $p^e | N$. For instance, 
         the denominator of $B_{1000}/1000$ is $2^4\cdot 3 \cdot 5^4\cdot 41\cdot 101 \cdot 251$. Thus the integrality of $<\tau>_g$ is obtained by the multiplication of the factor of the 
         (\ref{denominator1}) with a factor $(2g)!$.
         
         The denominator of $B_{2k}/k$ counts the numbers  of distinct  J-class map: $S^{m+4k-1}\to S^m$ as shown in (\ref{card}) \cite{Milnor}. The numerator of $B_{2k}/2k$ is related to differential topology such
         as characteristics. Indeed we have for $p=-1$, Euler characteristics $\zeta(1-2g) = (-1)^g B_{2g}/2g$ for $A_{p-1}$, and $( 1- 2^{2g - 1}) \frac{(-1)^g B_{2g}}{2g}$ for $D_l$ \cite{BrezinHikami5}.
         
         The asymptotic behavior for large $g$ of $B_{2g}$ is easily obtained by the formula,
         \be
         \zeta(2 g) = \frac{(2\pi)^{2g}}{ 2 (2g)!} |B_{2 g}|
         \ee
         Since $\zeta(2 g)= \sum \frac{1}{n^{2 g}} \sim 1$ (for $g\to \infty$), $|B_{2 g}| \sim \frac{2 (2 g)!}{(2 \pi)^{2 g}}$.
         
         The next leading term for $p\to \infty$ in (\ref{largep}) may be evaluated similar to the derivation of (\ref{largeplimit}). . From the lower order of $g$ in (\ref{A}) and (\ref{tauD}), the intersection numbers are expressed for $p\to \infty$ and a fixed $g$,
         \be
         <\tau>_{g} = \frac{|B_{2g}| p^g}{(2 g)!(2 g)} [ 1+  \frac{a_1(g)}{p} + \frac{a_2(g)}{p^2} + \cdots ]
         \ee
         where $a_n(g)$ is a polynomial of integer coefficients.
         
         Since the denominator of $<\tau>_g$ is given by the denominator of $B_{2g}/2g$ as shown above, it is interesting to note this implies the existence of a factor in the numerator of $<\tau>_g$. In the case of  $p=5$ ($A_4$ type), the denominator of $B_{10}/10$ is $2^2\cdot 3\cdot 11$, from the integrality of $a_1$ in (\ref{an}) (g=5), the factor $11$ should be in the numerator of $<\tau>_{g=5}$. Indeed the factor $2p+1$ gives 11 for $p=5$. The factor $(2p+1)$ i.e. $11$ exists in all $<\tau>_g$ and it cancels with  all denominator of $<\tau>_{g=5n}$, since $B_{5 g}/(5 g)$ has 11 in the denominator due to the formula of (\ref{denominator1}). The factor $(2^{10} 3^5 5^2)^n (\frac{3}{5})_n (\frac{4}{5})_n$ in (\ref{an}) gives the cancellation in part of  the denominator of $<\tau>_{5g}$ through the expression of Bernoulli number $B_{2g}$. 
         
         The another number of  series $c_{5n} = (\frac{4}{5})_n (\frac{1}{5})_n <\tau>_{5n}$, which is found to be integrality in \cite{Dubrovin} gives a factor $31,41,51,61,...$ due to $(\frac{1}{5})_n$. 
         Such prime numbers 31,41,61,.. appear in the denominator of $<\tau>_{5g}$ in higher genus $g$, and these prime numbers should be cancelled with Pochhammer
         symbol of $(\frac{1}{5})_n$. Thus  the integral numbers $a_n, b_n, c_n$ are consistent with the integrality of $<\tau>_{5g}$.
         
         The relation of the numerator of Bernoulli number $B_{2k}$ to differential topology $\Theta_{4k-1}$ is known as Milnor exotic 7 sphere. The order of $\Theta_{4k-1}$,
         card($\Theta_{4k-1}$) is given by the stable homotopy $\Pi_{4k-1}$ as \cite{Milnor,Mazur}
         \be\label{card}
         card (\Theta_{4k-1}) = 2^{2k-3}(2^{2k-1}-1) \cdot card(\Pi_{4k-1}) \cdot \frac{B_{2k}}{2k}
         \ee
         For $4k=2$, it becomes $28 = 2 (2^3-1) \cdot 240\cdot (\frac{1}{30})/4$. The numerator of $<\tau>_{g}$ has therefore a relation to differential topology and homotopy
         theory. Exotic 7-sphere is related $E_8$ singularity ($x^5+y^3 + z^2=0$) \cite{Milnor} and 28 different differential structures are described by algebraic equation such that $x_1^{2k-1} + x_2^3+ x_3^2 + 
         x_4^2 + x_5^2=0 (\Sigma(2k-1,3,2,2,2))$ (k=1,2,...,28) \cite{Brieskorn}.
         
         For $E_6$ singularity, the intersection numbers are evaluated by the ODE \cite{Dubrovin}, and the denominators of $\tau_{E_6}$ are not directly expressed by Bernoulli numbers , although it is closely related. The relation of Bernoulli number to mapping class group $\Gamma_g$ and characteristic classes has been discussed in \cite{Mislin}. The intersection numbers of half-integer $p$ spin curve  may have interesting further applications for topology and mapping class group.
         
         \vskip 3mm
         \section{n-point functions of $A_l$ types}
         \vskip 3mm
         
         The change of variable from $u$ to $y$ in (\ref{y}) is useful for the evaluation of higher point correlation functions $u(s_1,...,s_n)$.
         We have checked to obtain the known results for the integer values of $p=3,4,5$ of \cite{LiuXu,LiuVakilXu} by the Laurent expansion of $y$.
         
         \vskip 2mm
    {\bf{$\bullet$   two point functions for integer $p$}}
    \vskip 2mm
  
  The two point function $u(s_1,s_2)$ is written as, (for $p=\frac{1}{2}$, see (\ref{p1/2two})),
\ba\label{generalp}
    &&u(s_1,s_2) =  - 4 (\frac{s_1}{s_2})^{\frac{1}{p}} \oint \frac{dy_1 dy_2}{(2 i \pi)^2} (y_1 + \frac{s_1^{2+ \frac{2}{p}}}{y_1^3}) (y_2 + \frac{s_2^{2 + \frac{2}{p}}}{y_2^3}) \nonumber\\
    &&\times \frac{1}{y_1^4}
    \frac{ {\rm exp}[\sum_{i=1}^2 (-c) ( y_i^{2p}
     - \frac{p(2p+1)}{3} s_i^{2 + \frac{2}{p}} y_i^{2p-4}  + \frac{(2p+1) p (2p-1)(p-1)}{30} s_i^{4+ \frac{4}{p}} y_i^{2p-8} + \cdots)]}{[1 -\frac{s_1^{2+ \frac{2}{p}}}{y_1^4} - (\frac{s_1}{s_2})^{\frac{1}{p}} \frac{1}{y_1^2} ( y_2^2 - \frac{s_2^{2+ \frac{2}{p}}}{y_2^2})]^2 +\frac{ 4 s_1^{\frac{2}{p}}}{y_1^4} (s_1+ s_2)^2}\nonumber\\
    \ea
    For $p=\frac{1}{2}$, it becomes as (\ref{p1/2two}) with a slight difference of notation of $c$ (a factor 3 difference).

The selection rule of (\ref{RR}) is 
\be
2(g-1)(1+ \frac{1}{p}) + 2 = n_1+n_2 + \frac{1}{p}(j_1+j_2)
\ee
The two point function is expressed as
\be
u(s_1,s_2) = C\sum_{n_i,j_i} <\tau_{n_1,j_1}\tau_{n_2,j_2}> s_1^{n_1+ \frac{1}{p}(1+ j_1)} s_2^{n_2+ \frac{1}{p}(1+ j_2)}
  \ee
  where $C$ is a constant., which involves factors of gamma-function. This two point function and the  intersection numbers $<\tau_{n_1,j_1}\tau_{n_2,j_2}>$ for Neveu-Schwarz punctures are evaluated in general $p$ in \cite{BrezinHikami13}. Here we reconsider two point functions under the formula by the integral $y$ in (\ref{generalp}), which may be  easily obtained in  more systematic ways for both  Neveu-Schawrz and Ramond punctures. The numerator and denominator of (\ref{generalp}) are expanded
  in the small $s_1$ and $s_2$. 
 
 As an example, the case
  $<\tau_{2,1}\tau_{2,1}>$ of $p=4$, we find
 \ba
 ({\rm ii}.5) &&u(s_1,s_2) = s_1^{\frac{5}{2}}s_2^{\frac{5}{2}} \int_0^\infty dy_1 dy_2 e^{-c(y_1^8 + y_2^8)}( \frac{1}{y_1^{13}}(-2816 y_2^3 + 7680 y_2^{11})\nonumber\\
 &&+ \frac{1}{y_1^{21}}( -28728 y_2^{11} + 27888 y_2^{19} - 2880 y_2^{27}))
 = \frac{11}{240} s_1^{\frac{5}{2}}s_2^{\frac{5}{2}} [\Gamma(\frac{1}{2})]^2
 \ea
  By multiplying a factor $\frac{1}{p}= \frac{1}{4}$, it leads to $<\tau_{2,1}\tau_{2,1}>= \frac{11}{960}$, which agrees with \cite{LiuXu}.
  
 We have checked $<\tau_{1,1}\tau_{3,1}>_{g=2}= \frac{11}{4320}$ for $p=3$, $<\tau_{1,1}\tau_{3,1}>_{g=2} =\frac{17}{1200}$ for $p=5$, which are genus $g=2$ cases.

         \vskip 2mm
    {\bf{$\bullet$   three point functions for integer $p$}}
    \vskip 2mm
     
    Three point function $u(s_1,s_2,s_3)$ is given for $A$ type by
    \ba\label{u3}
   && u(s_1,s_2,s_3) = - 8 (\frac{s_1}{s_3})^{\frac{1}{p}} \int \frac{1}{(2i \pi)^3} \prod_{i=1}^3 d y_i \prod_{i=1}^3  (y_i + \frac{s_i^{2+ \frac{2}{p}}}{y_i^3}) e^{- c \sum (y_i^{2p} + \cdots)}\nonumber\\
   && \times \frac{1}{y_1^4}  \frac{1}{1 - \frac{s_1^{2+ \frac{2}{p}}}{y_1^4} 
   - (\frac{s_1}{s_2})^{\frac{1}{p}} \frac{1}{y_1^2}(y_2^2- \frac{s_2^{2+ \frac{2}{p}}}{y_2^2}) 
   - \frac{2 i s_1^{\frac{1}{p}}}{y_1^2}(s_1+ s_2)} \nonumber\\
   && \times  \frac{1}{y_2^2} \frac{1}{1 - \frac{s_2^{2+ \frac{2}{p}}}{y_2^4} 
   - (\frac{s_2}{s_3})^{\frac{1}{p}} \frac{1}{y_2^2}(y_3^2- \frac{s_3^{2+ \frac{2}{p}}}{y_3^2}) 
   - \frac{2 i s_2^{\frac{1}{p}}}{y_2^2}(s_2+ s_3)} \nonumber\\
   && \times  \frac{1}{1 - \frac{s_1^{2+ \frac{2}{p}}}{y_1^4} 
   - (\frac{s_1}{s_3})^{\frac{1}{p}} \frac{1}{y_1^2}(y_3^2- \frac{s_3^{2+ \frac{2}{p}}}{y_3^2}) 
   + \frac{2 i s_1^{\frac{1}{p}}}{y_1^2}(s_1+ s_3)} 
\ea
The exponential term is expressed after the scaling of $s$ as (\ref{generalp}),
\be\label{exponent}
e^{-y_i^{2p} + \frac{p(2p+1)}{3} s_i^{2+ \frac{2}{p}} y_i^{2p-2} - \frac{(2p+1)p (2p-1)(p-1)}{30} s_i^{4+ \frac{4}{p}}y_i^{2p-8} + \cdots }
\ee
where the factor $c$ is absorbed in $y_i$, which follows the scaling to $s_i$  as $c s^{p+1}$.

          There are results of the intersection numbers of three-point for $p=3$ up to $g=2$  for different 6 intersection numbers\cite{LiuXu}.
         We evaluated these 6 cases   to verify the validity the integral representation of (\ref{u3}) based on the random matrix theory.

          For instance, the case of $<\tau_{1,1}\tau_{1,1}\tau_{3,0}>_{g=2}$ is evaluated as,
       \ba
       ({\rm iii}.5) &&u(s_1,s_2,s_3) = -8 c^{\frac{5}{3}} s_1^{\frac{5}{3}}s_2^{\frac{5}{3}}s_3^{\frac{10}{3}} \int \prod dy_i e^{-\sum y_i^6} \biggl( - 560 \frac{y_2 y_3^3}{y_1^{11}} - 80 \frac{y_2}{y_1^{11}y_3^3} \nonumber\\
       && + \frac{1}{y_1^{11}y_2^5}(2860 y_3^3 - 7700 y_3^9 + 1470 y_3^{15}) + \frac{1}{y_1^{11}y_2^{11}} (8520 y_3^9- 8400 y_3^{15} \nonumber\\
       &&+ 1225 y_3^{21})\biggr)
       \nonumber\\
       &&= \frac{29}{1440} s_1^{\frac{5}{3}}s_2^{\frac{5}{3}}s_3^{\frac{10}{3}} (\Gamma(\frac{1}{3}))^2\Gamma(\frac{2}{3})
       \ea
       This leads to $<\tau_{1,1}\tau_{1,1}\tau_{3,0}>_{g=2}= \frac{29}{2160}$ for $p=3$ by the normalization of $\frac{2}{3}$.
       
         We have correctly derived the 6 intersection numbers  of three punctures of $p=3$ in genus $g=2$ with a normalization constant $\frac{2}{3}$, which agrees the values of \cite{LiuXu}.

         Thus the method of Laurent expansion of $y$ works for the evaluation of the higher point correlation function, which is a generating function of the intersection number, and
         it provides a practical method for higher  correlation functions.
         
    
    \vskip 3mm
    \section{n-point functions of $p= \frac{1}{2}, - \frac{1}{2}, -2$, and $-3$}
    \vskip 2mm
     In this section, we consider the half integer $p=\frac{1}{2}, -\frac{1}{2}, -\frac{3}{2}$ and the negative integer $p=-2,-3$ cases.  Some of these one point functions have been discussd  in the previous articles \cite{BrezinHikami1,BrezinHikami2}.

       \vskip 2mm
    {\bf{  7-1: n-point function for $p=\frac{1}{2}$}}
    \vskip 2mm
          
      Since $\frac{1}{p}\sum j_i$  is integer in this case, we are able to
     include them in the integral part of $\sum n_i$. Then the spin component $j_i$ can be chosen as arbitrary value as mod $p$. Here we take the spin component $j_i$ as $- 1$. Other choice may be $j=0$ or $j=-\frac{1}{2}$. All these cases provide the shift of the integer $n$. From the continuation of Ramond spin component $j=p-1$, the choice of $j=-\frac{1}{2}$ may be naturally considered. However, 
     the difference between $j=-1$ and $j=-\frac{1}{2}$ leads the shift of the integer $n$, this choice may do  not cause a serious conclusion. For the integer $p$, the algebraic geometry has been studied \cite{Witten2}.    
     
      One point function $u(s)$ is given by (\ref{onepoint1/2}), which becomes by an integral of part,
     \be
     u(s) = - \frac{i}{6 c' s^{3/2}} \oint \frac{dy}{2i\pi} e^{c' s^{3/2}(3y - \frac{1}{y^3})}
     \ee
     and it becomes a series of $\sum_g a_g s^{6g-3}$.
       \vskip 2mm
    {\bf{$\bullet$ string equation for $p= \frac{1}{2}$ }}
    \vskip 2mm

     We discuss the string equation for $p=\frac{1}{2}$.  Although the usual string equation may be not applied for this case, analogous equation
    about a forgetting of a marked point  can be considered. 

        The selection rule for $p=\frac{1}{2}$  (\ref{RR}) for $s$ marked points is,
    \be
    6 g - 6 + s = \sum_{i=1}^s n_i + 2\sum_{i=1}^s j_i
    \ee
    
    Possible values of $j_i$ may be $0,-\frac{1}{2},-1$, for which the last term becomes integers.
     The term $s_i^{n_i + \frac{1+m_i}{p}}$ becomes $s_i^{n_i+ 2}$, $s_i^{n_i+ 1}$ and $s_1^{n_i}$, respectively.
    
    By taking $j_i=-1$, it becomes
    \be\label{RR1/2}
    6g - 6 + 3s = \sum_{i=1}^s n_i
    \ee
       The term of order $s_1^2 s_2^{6g-2}$  are derived from (\ref{generalp}), which is considered as a string equation, since $s^{\frac{1}{p}}=s^2$ for $p=\frac{1}{2}$. 
    \ba\label{stringu21/2}
    u(s_1,s_2) &=& -  \frac{4 s_1^2}{s_2^2} \oint \frac{dy_1 dy_2}{(2 i \pi)^2} \frac{1}{y_1^3} (y_2 \pm \frac{s_2^6}{y_2^3}) e^{c_1 (3 y_1) + c_2 (3y_2 - \frac{s_2^6}{y_2^3})}\nonumber\\
    &=& \sum_g a_g s_1^2 s_2^{6g-2} c_1^2 c_2^{4g-2}
    \ea
    The coefficient $a_g$ is denoted by the intersection numbers as $<\tau_{2,-1} \tau_{6g-2,-1}>$, where we defined $n$ of $\tau_{n,-1}$ as the power of $s^{n+ \frac{1+ j}{p}}$. $<\tau_{n,-1}>$ is a coefficient of $s^n$. This term of order $s_2^{6g-2}$ is  one point function in (\ref{onepoint1/2}).
    Thus we have a string equation,
    \be
    <\tau_{2,-1} \tau_{6g-2,-1}> = <\tau_{6g-3,-1}>
    \ee
   in which $\tau_{2,-1}$ operates on  $\tau_{6g-2,-1}$ for the change to $\tau_{6g-3,-1}$.   Note that the usual notation of $\tau_{n,j}$ is different from above as a shift $n\to n-2$, since we took $j_i=-1$.
     
   Three points function of $p=\frac{1}{2}$  is studied for $A_l$ case in \cite{BrezinHikami2}. From the selection rule of $j_i=-1$ ($i=1,2,3$), we have from (\ref{RR1/2}) ; 
   $ 6g + 3 = n_1+ n_2 + n_3 $ for $<\tau_{n_1,-1}\tau_{n_2,-1}\tau_{n_3,-1}>$. 
   
       The term of order $s_1^2$ is obtained from three point function as
    \ba
    &&u(s_1,s_2,s_3) = - \frac{s_1^2}{(s_2 s_3)^2}\oint \frac{dy_1}{2i\pi} \frac{1}{y_1^3} e^{3 c_1 y_1} \oint \frac{dy_2 dy_3}{(2 i \pi)^2} (y_2 \pm \frac{s_2^6}{y_2^3})
    (y_3 +\pm \frac{s_3^6}{y_3^3})\nonumber\\
    && e^{c_2(3y_2- \frac{s_2^6}{y_2^3}) + c_3 (3y_3 - \frac{s_3^6}{y_3^3})}
    \frac{1}{2 i (s_2+ s_3)
    + \frac{y_2^2}{s_2^2} - \frac{s_2^4}{y_2^2} - \frac{y_3^2}{s_3^2}+ \frac{s_3^4}{y_3^2}}
    \ea
    Since two point function $u(s_2,s_3)$ is written as
    \ba
   && u(s_2,s_3)= \frac{i}{s_2^2 s_3^2(s_2+ s_3)}\oint \frac{dy_2dy_3}{(2 i \pi)^2} (y_2 \pm \frac{s_2^6}{y_2^3})(y_3 \pm \frac{s_3^6}{y_3^3})
    e^{c_2 (3 y_2- \frac{s_2^6}{y_2^3}) + c_3 (3 y_3 - \frac{s_3^6}{y_3^3})}\nonumber\\
    &&\biggl(\frac{1}{\frac{y_2^2}{s_2^2}- \frac{s_2^4}{y_2^2}- \frac{y_3^2}{s_3^2} + \frac{s_3^4}{y_3^2}- 2 i (s_2+ s_3)} - c.c. \biggr)
    \ea
    Thus we have a string equation
    \be
   <\tau_{2,-1}\tau_{n_2,-1}\tau_{n_3,-1}>_g = <\tau_{n_2-1,-1}\tau_{n_3,-1}>_g  + <\tau_{n_2,-1} \tau_{n_3-1,-1}>_g
   \ee

    The difference between $A_l$ and $D_l$ is due to the sign ($\pm$1) in the representation of $y$ variable
   in (\ref{pm}), and the string equation of three points function for $A_l$ ia valid also for $D_l$ case.
   
         \vskip 2mm
    {\bf{$\bullet$ two point function and absence of dilaton equation for $p= \frac{1}{2}$}}
    \vskip 2mm
    
    For integer spin $p$, a dilaton field corresponds to $\tau_{1,0}$, which corresponds to $s^{1+ \frac{1}{p}}$. When $p=\frac{1}{2}$ is inserted into this term, 
    it becomes  $s^{3}$ term.
    
    We have found that there is no dilaton equation for $p=\frac{1}{2}$ \cite{BrezinHikami2}. The one marked point
    intersection number  $<\tau_{1,0}>_{g=1}$ corresponds to  $<\tau_{3,-1}>_{g=1}$ in our notation. However, we do not find  two point function
    \be
    <\tau_3\tau_{6g-3}>_g = 0  \times  <\tau_{6g-3}>_g
    \ee
    which means a vanishing coefficient. Above equation plays a role of a dilaton equation, but the coefficient becomes zero. 
     
    The two point function of $p=\frac{1}{2}$ is obtained by the expansions of small $s_1$ and $s_2$ in a genus expansion.
    For this purpose, we write the integral representation of (\ref{generalp}) as
    \ba\label{pm2}
    u(s_1,s_2) &=& -\frac{4}{ s_1^2 s_2^2} \oint \frac{dy_1dy_2}{(2 i \pi)^2} ( y_1+ \frac{s_1^6}{y_1^3})(y_2 + \frac{s_2^6}{y_2^3}) e^{c \sum ( y_i - \frac{s_i^6}{3 y_i^3})}\nonumber\\
    && \times \frac{1}{(\frac{1}{ s_1^2}(y_1^2- \frac{s_1^6}{y_1^2}) - \frac{1}{ s_2^2}(y_2^2- \frac{s_2^6}{y_2^2}))^2 + 4 (s_1+s_2)^2}
    \ea

    If we  expand in the inverse of $y_1$, we obtain
    \ba\label{p1/2two}
    u(s_1,s_2) &=& - 4 \frac{s_1^2}{s_2^2}\oint \frac{dy_1 dy_2}{(2 i \pi)^2} (y_1 +  \frac{s_1^6}{y_1^3})(y_2 + \frac{s_2^6}{y_2^3}) e^{c (y_1- \frac{s_1^6}{3 y_1^3}) +
    c (y_2- \frac{s_2^6}{3 y_2^3})}\nonumber\\
    &\times&\frac{1}{y_1^4}[ 1 + g + g^2 + g^3 + \cdots]
    \ea
    where 
    \ba
    g &=& 2 (  \frac{s_1^6}{y_1^4} + \frac{s_1^2 y_2^2}{s_2^2 y_1^2} - \frac{s_1^2 s_2^4}{y_1^2 y_2^2}) - (\frac{s_1^6}{y_1^4} + \frac{s_1^2 y_2^2}{s_2^2 y_1^2} - \frac{s_1^2 s_2^4}{y_1^2 y_2^2})^2\nonumber\\
    &-& \frac{ 4 s_1^4 (s_1+s_2)^2}{y_1^4}
    \ea
    By the contour integral at $y_1=0,y_2=0$, we have expanding the exponential term proportional to $c$,
    \ba\label{twopoint1/2}
    &&u(s_1,s_2) =  - [ \frac{1}{3} c^4 s_1^2 s_2^4 \nonumber\\
    &&+  c^8 ( \frac{1}{360}  s_1^8 s_2^4
    + \frac{1}{135}  s_1^7 s_2^5 + \frac{1}{180} s_1^6 s_2^{6} + \frac{1}{216}  s_1^4 s_2^{8}+ \frac{1}{1080}s_1^2 s_2^{10}) + O(c^{12})] \nonumber\\
    &&+ [ s_1\leftrightarrow s_2 ]
    \ea
    The expression should be  symmetric by adding the terms of $s_1\leftrightarrow s_2$. The terms of order $g=2$ becomes by this symmetrization,
    \be
    c^8 \frac{1}{3^3 \cdot 5} \biggl(( s_1^8 s_2^4 + s_1^4 s_2^8 + s_1^7 s_2^5 + s_1^5 s_2^7) + \frac{3}{2} s_1^6 s_2^6 + \frac{1}{8} (s_1^2 s_2^{10} + s_1^{10} s_2^2)\biggr)
    \ee
   The first term shows the string equation; $s_1^2 s_2^4 \to s_2^3$.  Note that there is no $s^3$ term in two point function, which
   is supposed to be present for the term for a dilaton equation. 
   
       From (\ref{RR}) for $p=\frac{1}{2}$, we have $6g= n_1+n_2$ for $s_1^{n_1}s_2^{n_2}$. The first etrm of (\ref{twopoint1/2})
   is genus one, and the term of $c^8$ is genus two, since $n_1+n_2= 12$. It is interesting to note that $n_1$ and $n_2$ are both even, or both odd, since $6g$ is an even number.
   The terms of $s_1^2 s_2^{10}$ is consistent with  a string equation of (\ref{stringu21/2}).

     It may be easy to obtain the higher order terms. The parameter $c$ is $c=  \frac{i}{4}$. Each term of two point function is order of
    $c^{4g}s_1^{n_1}s_2^{6g-n_1}$ for genus $g$, which agrees with the selection rule of  (\ref{RR}). The scaling relation between $c$ and $s$ can be seen in (\ref{cs}). After the rescaling of $u\to \frac{s}{2} u$, the exponent becomes $c s^{p+1}$. This leads to the scaling $c s^{\frac{3}{2}}$, and $u(s_1,s_2) \sim \sum_{n=2} \sum_{g=1}^\infty c^{4g}s_1^n s_2^{6g-n}$. (\ref{twopoint1/2}) is consistent with this behavior.
    
    There may be another way for the evaluation of the contour integral. For genus one case, the last factor of (\ref{pm2}) is approximated by
    \be
    \frac{1}{(\frac{y_1^2}{s_1^2} - \frac{y_2^2}{s_2^2})^2} = \frac{1}{(\frac{y_1}{s_1}- \frac{y_2}{s_2})^2 (\frac{y_1}{s_1} + \frac{y_2}{s_2})^2}
    \ee
    We take  residues at $y_1= \frac{s_1}{s_2}y_2$ and at $y_1= - \frac{s_1}{s_2} y_2$. Expanding the exponent in order $c^4$, we find the residue of $y_1= \frac{s_1}{s_2}y_2$ and $y_2=0$ as
    \be
    U(s_1,s_2) = \frac{c^4}{6 s_2^3} ( s_1^9 + 2 s_1^8 s_2 + s_1^7 s_2^2 - s_1^3 s_2^6 - 2 s_1^2 s_2^7 - s_1 s_2^8 )
    \ee
    and from the pole at $y_1= - \frac{s_1}{s_2} y_2$ and $y_2=0$ as
    \be
    u(s_1,s_2) = \frac{c^4}{6 s_2^3} ( - s_1^9 + 2 s_1^8 s_2 - s_1^7 s_2^2 + s_1^3 s_2^6 - 2 s_1^2 s_2^7 + s_1 s_2^8 )
    \ee
    Adding these two contributions, by noting the cancellation of terms, we have
    \be
    u(s_1,s_2) = \frac{2 c^4 s_1^8}{3 s_2^2} - \frac{2}{3} c^4 s_1^2 s_2^4
    \ee
    The term of $s_1^2 s_2^4$ agrees with (\ref{twopoint1/2}) in genus one. 
    
    One may evaluate the residue of (\ref{pm2}) by taking 8 poles of $y_1$. There are 8 poles, except a pole of numerator $y_1=0$,
    \be
    y_1 = \pm \frac{i s_1 s_2^2}{2 y_2}\pm \frac{s_1 y_2}{2 s_2} \pm \frac{s_1 \sqrt{ - s_2^6 \pm 4 i s_1 s_2^2 y_2^2 \pm 2 i s_2^3 y_2^2 + y_2^4}}{2 s_2 y_2}
    \ee
    These residues for 8 poles are simply expressed as
    \ba
    &&\oint \frac{dy_1 }{2 i \pi} \frac{(y_1 +  \frac{s_1^6}{y_1^3})(y_2 + \frac{s_2^6}{y_2^3})}
   {(\frac{1} { s_1^2}(y_1^2- \frac{s_1^6}{y_1^2}) - \frac{1}{ s_2^2}(y_2^2- \frac{s_2^6}{y_2^2}))^2 + 4 (s_1+s_2)^2}\nonumber\\
    &&= \pm i \frac{s_1^2 (s_2^6 +  y_2^4)}{8 (s_1+s_2) y_2^3}
    \ea
    Adding the contribution of 8 poles, there is a cancellation. Thus the contribution comes only from the pole of $y_1=y_2=0$.

    \vskip 2mm
    {\bf{$\bullet$  three-point function for $p= \frac{1}{2}$ }}
    \vskip 2mm
    We have three point function of $p=\frac{1}{2}$ by putting $p=\frac{1}{2}$ in (\ref{u3}),
    \ba\label{three}
    &&u(s_1,s_2,s_3) = - 8 \frac{s_1^2}{s_3^2} \oint \prod_{i=1}^3 \frac{d y_i}{2i\pi} \prod_{i=1}^3 (y_i + \frac{s_i^6}{y_i^3}) e^{c \sum (3y_i - \frac{s_i^6}{y_i^3})}\nonumber\\
    && \times (\frac{1}{y_1^4 y_2^2}) \frac{1}{1 - \frac{s_1^6}{y_1^4} - \frac{s_1^2}{s_2^2 y_1^2} (y_2^2 - \frac{s_2^6}{y_2^2}) - \frac{2 i s_1^2}{y_1^2} (s_1+ s_2)}\nonumber\\
    && \times \frac{1}{1 - \frac{s_2^6}{y_2^4} - \frac{s_2^2}{s_3^2 y_2^2}(y_3^2- \frac{s_3^6}{y_3^2}) - \frac{2 i s_2^2}{y_2^2}  (s_2+ s_3)}\nonumber\\
    &&\times \frac{1}{1 - \frac{s_1^6}{y_1^4} - \frac{s_1^2}{s_3^2 y_1^2} (y_3^2- \frac{s_3^6}{y_3^2}) + \frac{2 i s_1^2}{y_1^2}(s_1+ s_3)}
    \ea
    where $c = - \frac{1}{6} i^{1/2}$. Expanding the denominators in the power series of $s_i$, we obtain  by the evaluation of residues,
    \ba\label{A3}
    &&u(s_1,s_2,s_3) = - 8\frac{s_1^2}{s_3^2} \biggl(  \frac{27}{4} c^4 s_3^6 + i \frac{81}{4} c^6( s_1^2 s_2 s_3^6 + 3 s_2^3 s_3^6 + 3 s_2^2 s_3^7 ) + O(c^8)\biggr)\nonumber\\
    \ea
    These terms are order of $c^{4g+2}s^{6g+3}$ according to the scaling relation between $c$ and $s$ as $c s^{3/2}$.  After the contour integration of $y_1$, the remaining term is odd for the exchange of $(s_2,y_2) \leftrightarrow (s_3,y_3)$.     For this order of $c^4$, the contour integral of $y_1$ is factorized as
    \ba\label{parity}
    U(s_1,s_2,s_3)&=&- 8 s_1^2 \oint \frac{dy_1}{2i \pi} \frac{y_1}{y_1^4} e^{c (3y_1- \frac{s_1^6}{y_1^3})} \oint \frac{dy_2 dy_3}{(2 i \pi)^2} (y_2+\frac{s_2^6}{y_2^3})(y_3+ \frac{s_3^6}{y_3^3})
    \nonumber\\
    &\times&\frac{e^{c( 3y_2- \frac{s_2^6}{y_2^3})+ c(3y_3- \frac{s_3^6}{y_3^3})}}{s_3^2 y_2^2 - \frac{s_2^6 s_3^2}{y_2^2} - s_2^2 y_3^2 + \frac{s_2^2 s_3^6}{y_3^2} - 2 i s_2^2s_3^2 (s_2+s_3)}
   \ea
   The denominator becomes $(s_3^2y_2^2- s_2^2 y_3^2)^{-1}$ in this order, and it is odd for the exchange $(s_2,y_2)\leftrightarrow(s_3,y_3)$.
    Therefore the first term  of order $c^4$ is vanishing by adding the symmetric terms of $s_2\leftrightarrow s_3$. Absence of order $c^4$ term agrees with (\ref{RR}), which implies the order of $c^{4g+2}s^{6g+3}$ for a genus $g$ term.
    
     The term of order $c^6$ in (\ref{A3}) contains $i=(-1)^{1/2}$, which comes from the term of $i (s_i+s_j)$ in the denominator of (\ref{three}).  
      Note that $c = - \frac{1}{6} i^{3/2}$, $i c^6$ is real number.  
      The term of order $c^6$ is also vanishing by the following reason. We have expanded  the large $y_1 > y_2 > y_3$. We have to consider also the case
      of $y_3 > y_2$. For this case, we  exchange $y_2 \leftrightarrow y_3$ as well as $s_2\leftrightarrow s_3$ with the complex conjugate of the term $- 2 i (s_2+ s_3)$, and this leads to
      the opposite sign of the term of $c^6$ with the exchange $s_2\leftrightarrow s_3$. Thus after adding this symmetric term, we obtain the cancellation of (\ref{A3}) and vanishing result of $c^6$ term. The fact that
      there is no three point function suggests the punctures are pairwise, and the vanishing three point function gives the simple structure of the spin $p= \frac{1}{2}$ case, which may be related to the simplification for Faber conjecture \cite{Faber}  as discussed in \cite{BrezinHikami2,Shadrin3}.

       \vskip 3mm
    {\bf{$\bullet$  four-point function for $p= \frac{1}{2}$}}
    \vskip 2mm
    The selection rule for four point function is $6 g + 6 = \sum_{i=1}^4 n_i$.
    The connected four point function $u(s_1,s_2,s_3,s_4)$ is expressed as in \cite{BrezinHikami2} with the same change of variables $u_i \to \frac{i}{2} (y_i^2- \frac{1}{y_i^2})$,
    by assuming order of smallness $s_1< s_2 < s_3 < s_4$,
    \ba\label{four}
    &&u(s_1,s_2,s_3,s_4) = - 16 \frac{s_1^2}{s_4^2} \oint \prod_{i=1}^4 (y_i + \frac{s_i^6}{y_i^3})e^{c \sum_{i=1}^4 (3 y_i - \frac{s_i^6}{y_i^3})}
    \nonumber\\
    &\times& \frac{1}{y_1^2 - \frac{s_1^6}{y_1^2}- \frac{s_1^2}{s_2^2}(y_2^2 - \frac{s_2^6}{y_2^2}) - 2 i s_1^2 (s_1+ s_2)}
    \frac{1}{y_2^2 - \frac{s_2^6}{y_2^2}- \frac{s_2^2}{s_3^2}(y_3^2 - \frac{s_3^6}{y_3^2}) - 2 i s_2^2 (s_2+ s_3)}\nonumber\\
    &\times& \frac{1}{y_3^2 - \frac{s_3^6}{y_3^2}- \frac{s_3^2}{s_4^2}(y_4^2 - \frac{s_4^6}{y_4^2}) - 2 i s_3^2 (s_3+ s_4)}
     \frac{1}{y_1^2 - \frac{s_1^6}{y_1^2}- \frac{s_1^2}{s_4^2}(y_4^2 - \frac{s_4^6}{y_4^2}) + 2 i s_1^2 (s_1+ s_4)}\nonumber\\
    \ea
    Since the four point function is cyclic about $y_i$ ($i=1,...,4$), we take $y_1$ and $y_3$ are large, and expansion of $1/y_1$ and $1/y_3$ in (\ref{four}).
    Similar to the three point function, we find that the contour integrals of (\ref{four}) are vanishing for the terms of order $c^i$ ($i=1,...,7$). The non-vanishing term
    appears at the order $c^8$ with a result of 
    \be\label{four1}
    u(s_1,s_2,s_3,s_4) = - 729 c^8 s_1^2 s_3^2 s_2^4 s_4^4 + O(c^{12} s^{18})
    \ee
    where $c= - \frac{1}{6} i^{1/2}$. 
    This is consistent with the selection rule of $6g + 6 = \sum_{i=1}^4 n_i$, where $n_i$ is a power of $s_i$. In this order, the genus $g$ is one, and it is consistent with the scaling
    of $c^{4g+4}s^{6 g+ 6} = c^{8} s^{12}$. The result of (\ref{four1}) is obtained for $y_1 > \{y_2,y_4\}$ and $y_3 > \{y_2,y_4\}$. The total expression is obtained by the permutation of the variables$\{s_1,...,s_4\}$.
    
    In this four point function, the numbers of punctures are even numbers (i.e. it is four). The punctures have the spin indices $j = -1$ as a Ramond puncture, and according to this spin indices, the selection rule $6g+6= \sum_{i=1}^4 n_i$ is obtained. Thus our explicit
    result of the four point function is consistent with the conjecture that Ramond punctures are pairwise. 
    
    The term of (\ref{four1}) is consistent with the successive string equations, which give the reduction of $s_1^2 s_3^2 s_2^4 s_4^4\to s_1^2 s_2^3 s_4^4\to s_3^2 s_4^4\to s_4^3$. Note that $<\tau_3>$ is non-vanishing one point intersection number.
    
      For $n$-point function ($n>1$), this Ramond puncture pairing conjecture states that if $n$ is odd, the $n$-point function is vanishing, and if $n$ is even, it is non-vanishing. For $n$-point function ($n \ge 6$),
    the contour integral becomes similar to $n=4$. Assuming $y_{j}$ ($j$ is odd) is large, and expand in the inverse power of $y_j$, we obtain the $n$-point function 
    $u(s_1,...,s_n)$ from the evaluation of the residues of $y_i$ at $y_i=0$. Through these residual calculations, we find the selection rule of (\ref{RR})  for $p= \frac{1}{2}$  is valid with the Ramond values of spin component
    $j=-1$, and Ramond punctures ($j= - 1$) appear  as pair-wise form.
    \vskip 3mm



        \vskip 2mm
    {\bf{$\bullet$ continuation from integer $p$ to $p=\frac{1}{2}$}} 
    \vskip 2mm
 For the continuation to $p= \frac{1}{2}$, the spin component $j$ is assumed to be (-1) \cite{BrezinHikami1,BrezinHikami2}. From the selection rule of (\ref{RR}) , we have for $p=\frac{1}{2}$ 
 \be
 6g-6+3= \sum_{i=1}^3 n_i + 2\sum_{i=1}^3 j_i
  \ee
 
  For general $p$, the three point function $<\tau_{0,0}\tau_{0,0}\tau_{0,p-2}>t_{0,0}t_{0,0}t_{0,p-2}$ is genus zero term, which corresponds to $s_1^{\frac{1}{p}}s_2^{\frac{1}{p}}
  s_3^{\frac{p-1}{p}}$ term. Putting $p=\frac{1}{2}$, this term becomes $s_1^2 s_2^2 s_3^{-1}$. This term is obtained from the contour integral in (\ref{three}) by expanding the second denominator about $2i s_2^2 s_3/y_2^2$ as
  \be
  u(s_1,s_2,s_3)_{g=0} = -8 s_1^2 s_2^2 s_3^{-1} \oint \prod_{i=1}^3 \frac{dy_i}{2i \pi} \frac{y_3^3}{y_1^3y_2^3}  e^{c \sum y_i}
  \ee
  However the contour integral of $y_3$ gives zero around $y_3=0$. Thus we have no 3-point function of $p=\frac{1}{2}$ at genus zero. This result is consistent with the conjecture of the punctures of Ramond type, which should be paired.
  
   The string equation of $p= \frac{1}{2}$ has been found in \cite{BrezinHikami2}. It is transform from $t_{0,0}\to t_2$, $(t_2= s^2)$. 
   
   For $u(s)$ is given by (\ref{onepoint1/2}). 
   \be\label{u1}
   u(s) = \frac{c^2}{48} s^3 - \frac{1}{5!}(\frac{1}{48})^2 c^6 s^9 + \cdots
   \ee
   This expression is continuation of $u(s)$ for the integer $p$, with $p= \frac{1}{2}$, since for general $p$, it is given by  \cite{BrezinHikami3} as
   \be\label{u1p}
   u(s) = \frac{1}{s^{1+ \frac{1}{p}}}[ \Gamma(1+ \frac{1}{p}) - \frac{p-1}{24} y \Gamma(1- \frac{1}{p}) + \frac{(p-1)(p-3)(1+ 2p)}{5! 4^2 3}y^2 \Gamma(1- \frac{3}{p}) + \cdots]
   \ee
   with $y= s^{2+ \frac{2}{p}}$. There appears a divergence in the limit of $p=\frac{1}{2}$ for $\Gamma$-function, and with the normalization of this factor, (\ref{u1p}) becomes equivalent to (\ref{u1}) as noted in Eq.(18) of \cite{BrezinHikami2}.
   
  From the results of (\ref{twopoint1/2}), $u(s_1,s_2)$ is expressed as
  \ba\label{u2p1/2}
  &&u(s_1,s_2) = - \frac{1}{3}c^4 s_1^2 s_2^4 + c^8 ( \frac{1}{1080} s_1^2 s_2^{10} + \frac{1}{216} s_1^4 s_2^8 + \frac{1}{180} s_1^6 s_2^6 + \frac{1}{135} s_1^7 s_2^5 \nonumber\\
  &&+ \frac{1}{360} s_1^8 s_2^4)
  \nonumber\\
  && - c^{12} ( \frac{s_1^2 s_2^{16}}{3265920} + \frac{s_1^4 s_2^{14}}{204120} + \frac{7 s_1^6 s_2^{12}}{291600} + \frac{s_1^7 s_2^{11}}{48600} + \frac{313 s_1^8 s_2^{10}}{8164800}
  + \frac{s_1^9 s_2^9}{22680} \nonumber\\
  &&+ \frac{443 s_1^{10}s_2^8}{16329600}
  + \frac{11 s_1^{11}s_2^7}{510300} + \frac{169 s_1^{12} s_2^6}{8164800} + \frac{19 s_1^{13}s_2^5}{2041200} + \frac{s_1^{14}s_2^4}{583200}) + O(c^{16})\nonumber\\
  \ea
  This expression is interpreted as a continuation of $p \to \frac{1}{2}$. $s_2^4$ corresponds to $s_2^{\frac{2}{p}}$ with $p=\frac{1}{2}$. Since $s_2^{\frac{2}{p}}$ is $t_{0,1}$, 
  $s^4$ corresponds to $t_{0,1}$. The term $s^5$ corresponds to $t_{1,1}$.
  
  The intersection number  for $s_1^{\frac{2}{p}}s_2^{4+ \frac{2}{p}}$ is given
  \be\label{48}
  <\tau_{0,1}\tau_{4,1}>_{g=2} = \frac{(p-1)(p+2)(p-2)}{2880 p} \to \frac{1}{768}
  \ee
  This expressesion agrees for the value of \cite{LiuXu,LiuVakilXu}, $\frac{1}{320}$ and $\frac{7}{1200}$ for $p=4$ and $p=5$.
  (\ref{48}) corresponds to $s_1^{\frac{2}{p}}s_2^{4+\frac{2}{p}}\to s_1^4 s_2^8$ for $p=\frac{1}{2}$.
  The term, which gives $s_1^4 s_2^8$ in the limit $p\to \frac{1}{2}$, may exists also from another term $<\tau_{2,0}\tau_{2,2}>$ with $s_1^{2 + \frac{1}{p}}s_2^{2+ \frac{3}{p}}$.
  Therefore, $s_1^4 s_2^8$ is obtained from the limit of different terms of general $p$. The term $s_1^{2+ \frac{1}{p}}s_2^{2 + \frac{3}{p}}$ is as
  \be
  u(s_1,s_2) = - \frac{1}{576} (p+1)(p- 1) \Gamma(1-\frac{1}{p})\Gamma(1- \frac{3}{p}) s_1^{2+ \frac{2}{p}}s_2^{2+ \frac{1}{p}}
  \ee
  which by the change $s_1\leftrightarrow s_2$, it becomes $s_1^4s_2^8$ for $p\to \frac{1}{2}$.
  
  For $s_1^6s_2^6$ term in (\ref{u2p1/2}), we have
  \be\label{66}
  <\tau_{0,2}\tau_{4,0}>_{g=2} = \frac{(p-1)(p-3)(2p+11)}{5760 p}\to \frac{1}{192}
  \ee
  which corresponds to $s_1^{\frac{3}{p}}s_2^{4+ \frac{1}{p}}\to s_1^6 s_2^6$ for $p= \frac{1}{2}$. The values of these intersection numbers agree with the intersection
  numbers from (\ref{u2p1/2}).
  
  For $<\tau_{2,1}\tau_{2,1}>_{g=2}$ corresponds to $s_1^{2+\frac{2}{p}}s_2^{2+ \frac{2}{p}}\to s_1^6 s_2^6$, which degenerates to (\ref{66}). Thus the continuation to $p=\frac{1}{2}$ 
  is not unique.
  
  The term $s^{\frac{2}{p}}$ and $s^{2+ \frac{1}{p}}$ become same $s^4$ in the limit $p\to \frac{1}{2}$. The terms $s^{\frac{2}{p}}$, $s^{2+ \frac{2}{p}}$ and $s^{4+ \frac{1}{p}}$ become
  $s^6$ in the limit of $p\to \frac{1}{2}$.

    \vskip 3mm
           \vskip 2mm
    {\bf{7-2: n-point function of  $p= - \frac{1}{2}$}} 
    \vskip 2mm

  For $p= -\frac{1}{2}$, the selection rule becomes for $k$-marked points
  \be\label{-1/2}
  2g - 2 + k = -\sum_{i=1}^k n_i
  \ee
  where we take $j_i=-1$.
  This requires the negative power $n_i$ of $s$ for higher genus $g$, which leads to the inverse power of $s_i$ in $m$-point function $u(s_1,...,s_m)$.
  As discussed in \cite{BrezinHikami1,BrezinHikami2}, one point function $u(s)$ is a contour integral,
  \ba\label{one-1/2}
  u(s) &=& s \oint \frac{dy}{2 i \pi} (y + \frac{1}{s^2 y^3}) e^{- \frac{c}{y}}\nonumber\\
  &=& \frac{1}{2} c^2 s
  \ea
  This term belongs to genus $g=0$ order, which leads to $n_1=1$ from (\ref{-1/2}) as $s^{n_1}$. The coefficient $c^2$ is given by the scaling $(cs^{p+1})^2= c^2 s$.
  From the expansion of the intersection numbers in (\ref{A}), the higher order terms are vanishing for $p= -\frac{1}{2}$
   by the existent factor $(2p + 1)$  for $g > 1$.  In the genus $g=1$ order, the intersection number $<\tau_{1,0}> = \frac{p-1}{24}$ in (\ref{us})  for the integer $p$ suggests the nonvanishing
    value for $p=-\frac{1}{2}$.  This order is given by the second term of the integrand. By the change of $y\to \frac{1}{t}$, (\ref{one-1/2}) is written as
  \be\label{p/2one}
  u(s) = s \int \frac{dt}{2i \pi} (-\frac{1}{t^2}) (\frac{1}{t} + \frac{t^3}{s^2}) e^{-c t}
  \ee
The second term of the integrand becomes by the integration as $ - \frac{1}{c^2 s}$ $(c= (\frac{1}{4})^p= 2)$, which is a continuation of $\frac{p-1}{24}= -\frac{1}{16}$.  There is a factor $\Gamma(3)=2$ in (\ref{us}). Thus genus $g=1$ term is consistent with (\ref{us}).
  
  Two point function of (\Ref{generalp}) is written for $p= -\frac{1}{2}$ as
  \ba\label{p-1/2two}
  u(s_1,s_2) &=& - 4 (\frac{s_2}{s_1})^2 \oint \frac{dy_1 dy_2}{(2i\pi)^2} \prod_{i=1}^2 (y_i+ \frac{1}{s_i^2 y_i^3}) e^{- c\sum \frac{1}{y_i}}\nonumber\\
  && \frac{1}{y_1^4}\frac{1}{(1- \frac{1}{s_1^2y_1^4} - (\frac{s_2}{s_1})^2 \frac{1}{y_1^2}(y_2^2- \frac{1}{s_2^2y_2^2}))^2 + \frac{4}{s_1^4 y_1^4}(s_1+s_2)^2}
  \ea
  
  The contour integral is evaluated for the poles of $y_1=y_2=0$. By the expansion of the exponential term as $\sum (-c (\frac{1}{y_1} + \frac{1}{y_2}))^k/k! $,
  we obtain
  \ba\label{p-1/2}
  &&u(s_1,s_2) = - 4 s_2^2 ( \frac{1}{4}  c^4  - \frac{c^8}{2 \cdot 6!} (s_1^2 + 8 s_1 s_2 + 12 s_2^2)\nonumber\\
  && + \frac{c^{12}}{2\cdot 10!}(s_1^4 + 24 s_1^3 s_2+ 161 s_1^2 s_2^2 + 384 s_1 s_2^3+ 288 s_2^4)\nonumber\\
  && - \frac{c^{16}}{2\cdot 14!} (s_1^6 + 48 s_1^5 s_2 + 715 s_1^4 s_2^2 + 4528 s_1^3 s_2^3 + 13441 s_1^2 s_2^4 + 18304 s_1 s_2^5\nonumber\\
  && + 9152 s_2^6) + O(c^{20})) + (s_1 \leftrightarrow s_2)
  \ea
  Each term is written as  $c^{4 g} s_1^{|n_1|} s_2^{|n_2|}$, $n_1+ n_2 = 2g$. For $p=- \frac{1}{2}$, the selection rule becomes for two point function as ($m_i= -1$)
  \be
  - 2 g = n_1+ n_2
  \ee
  Thus $n_1,n_2$ are negative integers for $s_1^{-n_1} s_2^{-n_2}$. The expansion of (\ref{p-1/2}) is similar to the weak coupling expansion of $p=-2$ case.
  
  By the change of variable $y_i= \frac{1}{t_i}$, ($i=1,2$), it becomes by $t_i \to t_i/c$,
  \ba\label{2-1/2}
  u(s_1,s_2) &=& - 4 (\frac{s_2}{s_1})^2 \oint \frac{dt_1 dt_2}{(2 i \pi)^2} (t_1 + \frac{t_1^5}{c^4 s_1^2})e^{- t_1} (\frac{1}{t_2^3} + \frac{t_2}{c^4 s_2^2}) e^{- t_2}\nonumber\\
  && \frac{1}{(1 - \frac{t_1^4}{c^4 s_1^2} - (\frac{s_2}{s_1})^2 t_1^2 (\frac{1}{t_2^2} - \frac{t_2^2}{c^4 s_2^2}))^2 + \frac{4 t_1^4}{c^4 s_1^4}(s_1+s_2)^2}
  \ea
  From the selection rule of (\ref{-1/2}), in the case two point function $n=2$, we have 
  \be 
  2g = - (n_1+ n_2)
  \ee
  for $u(s_1,s_2) = \sum c_{n_1,n_2} s_1^{n_1} s_2^{n_2}$. The values of $n_1$ and $n_2$ can be both positive and negative integers.
  The expansion of (\ref{2-1/2}) is
  \ba
  u(s_1,s_2) &=&-4 (\frac{s_2}{s_1})^2 \oint \frac{dt_1 dt_2}{(2i\pi)^2} e^{-t_1-t_2}  \frac{t_1 t_2}{(t_2^2- \frac{s_2^2}{s_1^2} t_1^2 )^2}
  \biggl( 1 + \frac{t_1^4}{c^4 s_1^2} + \frac{t_2^4}{c^4 s_2^2}  \nonumber\\
  && + 2 \frac{t_1^2}{c^4 s_1^2} \frac{t_1^2- t_2^2}{1- \frac{s_2^2}{s_1^2} \frac{t_1^2}{t_2^2}} - 4 \frac{(s_1+s_2)^2}{c^4 s_1^4} \frac{t_1^4}{(1- \frac{s_2^2}{s_1^2} \frac{t_1^2}{t_2^2})^2} + O(\frac{1}{c^8})\biggr)
  \ea
  where the dependence of $c$ is obtained from the scaling to $s$ as a combination of 
  $c^2 s$. 
   \vskip 2mm
    {\bf{ 7-3: Strong coupling  of $p= - 2$}}
    \vskip 2mm

        \vskip 2mm
    {\bf{$\bullet$  $p= -2$ and unitary matrix model}} 
    \vskip 2mm
  It was pointed out that for $p=-2$ corresponds to unitary matrix model \cite{Mironov,BrezinHikami8}. The unitary matrix model  has two phases
  of the weak coupling and string coupling phases \cite{BG,GW}. 
     
    For one point function, we have for large $s$,
  \ba\label{coefficient}
  &&u(s) = \frac{i}{2} \oint \frac{dy}{2i \pi} (y + \frac{1}{y^3}) e^{\frac{c}{s} \frac{y^4}{(y^4 + 1)^2}}\nonumber\\
  &&= \frac{2}{s} + \frac{2}{ s^2} + \frac{2}{ s^3} + \frac{5 }{3 s^4} + \frac{7}{6 s^5} + \frac{7}{10 s^6} + \frac{11}{30 s^7} + \frac{11\cdot 13}{840 s^8} + \frac{11\cdot 13}{2016 s^{9}}\nonumber\\
  &&+ \frac{11\cdot 13\cdot 17}{2^5 3^4 5\cdot 7 s^{10}}+ \frac{13\cdot 17\cdot 19}{2^5 3^4 5^2\cdot 7 s^{11}} + \frac{13\cdot 17\cdot 19}{2^6 3^4 5^2\cdot 11 s^{12}}
  + \frac{17\cdot 19\cdot 23}{2^7 3^5 5^2 \cdot 11 s^{13}}
  + \cdots\nonumber\\
  \ea
  where $c= (\frac{1}{4})^p= 2^4$. The contour integral reduces to the  calculation of the residue at $y= (-1)^{1/4}$.
  This expansion is written by the modified Bessel function  as
  \ba\label{strong1}
  u(s) &=&
 = \frac{1}{s} \sum_{k=0}^\infty \frac{2^{2k+1} \Gamma(k+ \frac{1}{2})}{\Gamma(\frac{1}{2}) \Gamma(k+2)k!} \frac{1}{s^k}\nonumber\\
 &=& \frac{2}{s}F (\frac{1}{2},2; \frac{4}{s}) = \frac{2}{s} e^{\frac{2}{s}}[ I_0(\frac{2}{s})- I_1(\frac{2}{s})]
  \ea
  where $F(\alpha,\gamma; z)$ is a confluent hypergeometric function, and $I_0(z)$ is a modified Bessel function. The crossover to the weak coupling of small $s$ is at $s=2$, which corresponds to $s = \frac{1}{\Lambda}= 2$.
  The external source is $\Lambda$.
  The small $s$ expansion is given by (\ref{A}) through the intersection numbers of $p=-2$ \cite{BrezinHikami8}. We have
  \ba
  u(s) &=& \frac{1}{s}\oint \frac{du}{2i \pi} e^{- \frac{1}{u+s} + \frac{1}{u}}\nonumber\\
  &=& - \frac{1}{4 \pi \sqrt{s}}\int \frac{dx}{x^2} e^{- \frac{4 x^2}{1 + x^2 s}}\nonumber\\
  &=& - \frac{1}{4 \pi \sqrt{s}}\sum_{n=0}^\infty \frac{(2 n-1)!!}{2^n n!}\Gamma(n - \frac{1}{2}) ( \frac{s}{4})^n
  \ea
  By noting that $(2n-1)!!= 2^n \Gamma( n + \frac{1}{2})/\sqrt{\pi}$,  the last sum is written as
  \be
  u(s) = - \frac{1}{4 \pi^{\frac{3}{2}}\sqrt{s}} \sum_{n=0}^\infty \frac{\Gamma( n+ \frac{1}{2})\Gamma( n - \frac{1}{2})}{n!} (\frac{s}{4})^n
  \ee
  This weak coupling expansion of one point function has been derived and the comparison with the unitary matrix model has been discussed in \cite{BrezinHikami8}.
  
   The model $p=-2$ has naturally a logarithmic potential, which becomes equivalent to unitary $U(N)$ matrix model. 
   The strong coupling expansion in (\ref{strong1}) is evaluated by the character expansion  of $U(N)$ group, which character $\chi$ satisfies
   \be\label{character}
   \int dU \chi_r(A U) \chi_{r'} (U^\dagger A^\dagger) = \delta_{r r'} \frac{\chi_r(A A^\dagger)}{d_r}
   \ee
   The character of $U(N)$ is expressed by
   \be
   \chi_{n_1,n_2,...} = \frac{{\rm det}[ t_i^{n_j + N - j} ]}{{\rm det}[ t_i^{N-j}]}
   \ee
    where $t_i$ is eigenvalue of $U$. The dimension $d_r$ is equal to $d_r= \chi_r(1)$. For the eigenvalue 1, ($t_i=1$), 
    \be
    d_{[2,0]}= \chi_{[2,0]}(1) = \frac{1}{2}(N^2+ N), \hskip 3mm  d_{[1,1]}= \chi_{[1,1]}(1) = \frac{1}{2} (N^2-N)
    \ee
    The integration of unitary matrix $U$ is written as
    \be
    \int dU {\rm tr} (A U ) {\rm tr} (B U^{\dagger}) = C_1 {\rm tr} (A B)
    \ee
    where $A$ and $B$ are arbitrary matrices,
    \ba
   && \int DU {\rm tr} (A_1 U){\rm tr} (A_2 U) {\rm tr} (B_1 U^{\dagger}) {\rm tr }(B_2 U^{\dagger}) = C_{1^2} [{\rm tr}(A_1B_1){\rm tr}(A_2 B_2)\nonumber\\
   &&+ {\rm tr} (A_1 B_2) {\rm tr}(A_2 B_1)] + C_2 [ {\rm tr}(A_1B_1A_2B_2) + {\rm tr}(A_1 B_2 A_2 B_1)]
    \ea
    Generalization of this equation for $n$-times trace is described by $C_{l_1,...,l_n}$.
    
  The coefficient of the single trace is $C_n= C_{n,0,...0}$, which are
  \ba\label{C_n}
  &&C_1= 1,\hskip 2mm C_2 = - \frac{1}{N^2-1}, \hskip 2mm
  C_3= \frac{4}{(N^2-1)(N^2-4)},\nonumber\\
  &&C_4= - \frac{30}{(N^2-1)(N^2-4) (N^2-9)}, \nonumber\\
  &&C_5 = \frac{336}{(N^2-1)(N^2-4)(N^2-9)(N^2-16)}
  \ea
  The general formula of $C_n$ is \cite{Samuel}
  \ba\label{generalN}
  C_n &=& (-1)^n \frac{[(n-1)!]^3}{n} \frac{1}{(N^2-1)(N^2- 4)(N^2-9)\cdots (N^2- (n-1)^2)} \nonumber\\
  &\times& \sum_{q=0}^{n-1} \frac{1}{q! q! (n-q-1)! (n-q-1)!}
  \ea
  The coefficient of $1/s^n$ in (\ref{coefficient}) is $ 2 C_n$, and agrees with $N=0$ \cite{BrezinHikami8,Samuel}.
  
  It is interesting to note that when we change the measure $(y+ \frac{1}{y^3})$ to $(y- \frac{1}{y^3})$, i.e. D-type, one point function is vanishing i.e. $u(s)=0$. The replacement of this measure $y+1/y^3$ by $y$, the values of $u(s)$ becomes half of (\ref{coefficient}).
  
   For the derivation of the $N$ dependence, we make a  change of variable $u$ to $z$ as
   \be\label{changeuz}
   \frac{u-1}{u+1} =  e^{-z}, \hskip 3mm u = \frac{1+ e^{-z}}{1- e^{-z}}
   \ee
   and 
   \be
   du = - 2 \frac{e^{-z}}{(1 - e^{-z})^2}
   \ee
   Thus we obtain
   \ba\label{strongp=-2}
   U(s) &=& \frac{N}{2}\int du e^{\frac{4 c}{s (u^2-1)}} (\frac{u-1}{u+1})^N \nonumber\\
   &=&-  N\int_0^\infty dz \sum_{n=0}^\infty  \frac{1}{n!} e^{-Nz} [ \frac{c}{s} e^{z} ( 1- e^{-z})^2 ]^n \frac{e^{-z}}{(1- e^{-z})^2}
   \ea
   The leading term ($c=0$) gives the Euler characteristics $\zeta(1- 2g)$ \cite{BrezinHikami3}, the next order is $\frac{c}{s }$, and the second order becomes
   \be
   N\frac{c^2}{2 s^2} \int_0^\infty dz e^z ( 1- e^{-z})^2 e^{-N z} = \frac{c^2}{s^2 } \frac{1}{N^2-1}
   \ee
   The next order is
   \ba
   &&-\frac{N}{6}(\frac{c}{s})^3 \int dz e^{2z} ( 1 - e ^{-z})^4 e^{- N z}\nonumber\\
   &&= - \frac{N}{6}(\frac{c}{s})^3 [ \frac{1}{N-2} - \frac{4}{N-1} + \frac{6}{N} - \frac{4}{N+1} + \frac{1}{N+2}]\nonumber\\
   &&= - 4N (\frac{c}{s})^3 \frac{1}{N(N^2-1)(N^2- 4)}
   \ea
    Extending these evaluations,  we find from (\ref{strongp=-2})
   \be\label{Il[1]}
   U(s) = \sum_{n=1}^\infty \frac{(2n-2)!}{n! s^n} \frac{1}{\prod_{l=1}^{n-1} (N^2- l^2)}
   \ee
   where $c=1$. The number $(2n-2)!/n!$ coincides with the numerator of $C_n$ in (\ref{C_n}). Thus we find that one point function $U(s)$, which is a single trace operator, is a generating function of  the coefficient $C_n$ in (\ref{generalN}).

   For two point function $U(s_1,s_2)$ with a logarithmic term is written as
   \ba
   &&U(s_1,s_2) =\int du_1 du_2 e^{\frac{4c}{s_1 (u_1^2-1)} + \frac{4 c}{s_2(u_2^2-1)}} (\frac{u_1+1}{u_1-1})^N
   (\frac{u_2+1}{u_2-1})^N \nonumber\\
   && \times \frac{1}{(s_2 u_2 - s_1 u_1 + (s_1+ s_2))(s_2 u_2 -s_1 u_1 - (s_1+ s_2))}
   \ea
   
   Using the change of variables in (\ref{changeuz}), one obtain
   \ba
   &&U(s_1,s_2) = s_1 s_2 \int dz_1 dz_2 e^{-N z_1 - N z_2}  \frac{e^{-z_1}}{(1- e^{-z_1})^2} \frac{e^{-z_2}}{(1- e^{-z_2})^2} \nonumber\\
   && \times \frac{1}{s_2 ( 1 + \frac{2 e^{-z_2}}{1 - e^{-z_2}}) - s_1 ( 1+ \frac{2 e^{-z_1}}{1- e^{- z_1}}) + (s_1+s_2)}\nonumber\\
   && \times \frac{1}{s_2 ( 1 + \frac{2 e^{-z_2}}{1 - e^{-z_2}}) - s_1 ( 1+ \frac{2 e^{-z_1}}{1- e^{- z_1}}) -( s_1+s_2)}\nonumber\\
    && \times e^{\frac{c}{s_1}e^{z_1}(1- e^{-z_1})^2 + \frac{c}{s_2}e^{z_2} (1 - e^{-z_2})^2}
    \ea
   Expanding the two denominators for $s_2 > s_1$,
    \be
    \frac{1}{4 s_2^2} \sum_{m=0}^\infty \sum_{l=0}^\infty (\frac{1 - e^{-z_2}}{1- e^{-z_1}} e^{-z_1})^m (\frac{1- e^{-z_2}}{1- e^{-z_1}} e^{z_2})^l (\frac{s_1}{s_2})^{m+l} 
    e^{z_2} (1- e^{-z_2})^2 
    \ee
  with a measure of $du_i= e^{-z_i}/(1- e^{-z_i})^2 dz_i$, we obtain the strong coupling expansion of $1/s_1^{n_1} s_2^{n_2}$,
  \ba\label{twopointlarges}
  &&U(s_1,s_2) =(s_1 s_2) \int dz_1 dz_2  e^{- N z_1 - N z_2} \frac{e^{-z_1}}{(1- e^{-z_1})^2} \frac{e^{-z_2}}{(1- e^{-z_2})^2} \nonumber\\
    &&\times  \frac{1}{4 s_2^2} \sum_{m=0}^\infty \sum_{l=0}^\infty (\frac{1 - e^{-z_2}}{1- e^{-z_1}} e^{-z_1})^m (\frac{1- e^{-z_2}}{1- e^{-z_1}} e^{z_2})^l (\frac{s_1}{s_2})^{m+l} 
    e^{z_2} (1- e^{-z_2})^2 \nonumber\\
    && \times e^{\frac{c}{s_1}e^{z_1}(1- e^{-z_1})^2 + \frac{c}{s_2}e^{z_2} (1 - e^{-z_2})^2}
 \ea
  The leading term is given by Euler characteristics $\zeta(1- 2g)$,
  \ba
   &&U(s_1,s_2) =\frac{1}{4 s_2^2} \int dz_1 dz_2  e^{- N z_1 - N z_2} \frac{e^{-z_1}}{(1- e^{-z_1})^2}  \nonumber\\
    &&= \frac{1}{4 s_2^2 N^2}\sum_{n=0} \frac{1}{N^{2n}}(-1)^{n-1} \frac{B_n}{2n}\nonumber\\
    &&= \frac{1}{4 s_2^2 N^2} \sum_{n} \frac{1}{N^{2n}} \zeta(1- 2n)
    \ea
     where we have to count $s_1\leftrightarrow s_2$, and $B_n$ is Bernoulli number $B_1= \frac{1}{6}, B_2= \frac{1}{30},B_3= \frac{1}{42},...$. Note this term has no $s_1$ dependence.
     
     The next term in the strong coupling is the term of order $1/s_1 s_2$ as we will see later for $N=0$ in (\ref{strongexpansion}). 
     In the strong coupling region, $U(s_1,s_2)$ is expanded in the inverse power of $s_1$ and $s_2$. The pole of $(1- e^{-z_1})$ in (\ref{twopointlarges}) should be cancelled with the
     expansion of the exponent in the last term. Then, similar to one point function $U(s)$, two point function is evaluated in the form of characters of $U(N)$ as shown in \cite{BrezinHikami8}, which used a different method as shown here.

     We have from (\ref{twopointlarges}) two point function for $p=-2$,
     \ba\label{U2strong}
     U(s_1,s_2) &=& \frac{c}{4  s_2 N^2} + \frac{ c^2}{4 s_1 s_2} \frac{1}{N^2 (N^2-1)} +\frac{c^3}{s_1^2 s_2}\frac{1}{N^2 (N^2-1) (N^2-4)} \nonumber\\
     && + \frac{30 c^4}{4 s_1^3 s_2}\frac{1}{N^2(N^2-1)(N^2-4)(N^2-9)} + O(\frac{1}{ s_1^n s_2^2})
     \ea
   
   The coefficient $C_{1,n}$ is \cite{BrezinHikami8}
   \ba
   &&C_{1,1} = \frac{1}{N^2-1}, \hskip 3mm C_{1,2}= - \frac{12}{(N^2-1)(N^2-4)}\nonumber\\
    &&C_{1,3} = \frac{120}{(N^2-1)(N^2-4)(N^2-9)} \nonumber\\
    &&C_{1,4} = - \frac{1680}{(N^2-1)(N^2-4)(N^2-9)(N^2-16)}
   \ea
   The expression of $U(s_1,s_2)$ of (\ref{twopointlarges}) in a strong coupling agrees with the coefficients of $C_{n_1,n_2}$.

This shows that
the $n$ point function $U(s_1,...,s_n)$ of $p=-2$ case in the inverse $s$ expansion is a generating function of the coefficient $C_{l_1,...,l_n}$ of 
unitary integral of the multi-trace products.

 We now turn to the change of variable $y$.  We here use the expression of (\ref{generalp}) for $p=-2$.
  It becomes for two point function,
  \ba\label{strong}
  &&u(s_1,s_2) = -4 s_1 s_2 \oint \frac{dy_1 dy_2}{(2i \pi)^2} (\frac{1+ y_1^4}{y_1^3}) (\frac{1 + y_2^4}{y_2^3}) e^{- \frac{c}{s_1} \frac{y_1^4}{(1+ y_1^4)^2}
  - \frac{c}{s_2} \frac{y_2^4}{(1+ y_2^4)^2}}\nonumber\\
  && \times \frac{1}{(s_1(y_1^2- \frac{1}{y_1^2}) - s_2 (y_2^2 - \frac{1}{y_2}))^2 + 4 (s_1+ s_2)^2}
  \ea
  For the large $s$ expansion, the expansion parameter is coupled to $c$ as $\frac{c}{s}$ by a scaling. The exponent of the integrand has a pole at $1+ y_i^4 = 0$, 
  which leads to $y_i = \pm (-1)^{\frac{1}{4}}, y_i= \pm (-1)^{\frac{3}{4}}$. 
  This pole appears for $p \le - \frac{3}{2}$, since the term of exponent is $(y^2 + \frac{1}{y^2})^{2p+ 2}$. Taking the pair of poles $y_1 = \pm (-1)^{\frac{1}{4}}, y_2= \pm (-1)^{\frac{1}{4}}$ or $y_1= \pm (-1)^{\frac{3}{4}},
  y_2= \pm (-1)^{\frac{3}{4}}$, two point function in the large $s$  is evaluated with a contour integral as,
  \ba\label{strongexpansion}
  &&u(s_1,s_2) = \frac{c^2}{2^6} \frac{1}{s_1 s_2} - \frac{c^3}{2^{10}} (\frac{1}{s_1 s_2^2} + \frac{1}{s_1^2 s_2}) 
   + \frac{c^4}{3\cdot 2^{15}}(   \frac{5}{s_1 s_2^3}  + \frac{7}{s_1^2 s_2^2} + \frac{5}{s_1^3 s_2})\nonumber\\
  &&- \frac{c^5}{2^{20} \cdot 3} (  \frac{7}{s_1 s_2^4} + \frac{13}{s_1^2 s_2^3} + \frac{13}{s_1^3 s_2^2}
   + \frac{7}{s_1^4 s_2}) \nonumber\\
   && + \frac{c^6}{2^{24} \cdot 15}(\frac{21}{s_1 s_2^5}+ \frac{49}{s_1^2 s_2^4} + \frac{66}{s_1^3 s_2^3}
   + \frac{49}{s_1^4 s_2^2}+ \frac{21}{s_1^5 s_2})\nonumber\\
   && - \frac{c^7}{2^{28}\cdot 15} 
   ( \frac{11}{s_1s_2^6} + \frac{31}{s_1^2 s_2^5} + \frac{53}{s_1^3 s_2^4} + \frac{53}{s_1^4 s_2^3} + \frac{31}{s_1^5 s_2^2} + \frac{11}{s_1^6 s_2})
   + O(c^8)
   \ea
  where $c= (\frac{1}{4})^p = 2^4$. This result is consistent with \cite{BrezinHikami8}, where  there is a factor $N$, which comes from the logarithmic 
  potential. Above expansion coincides with the result of $N=0$ in \cite{BrezinHikami8}, since we do not take the logarithmic potential in (\ref{strong}).   
    From this derivation, it can be found that the large $s$ region is  related to the pole at the solution of  $y^4 + 1 = 0$. 
  This pole exists for $p = -2$, while this pole does not exists for the positive integer $p$. It may be interesting to discuss $p=-3/2$ and $p = -3$ cases, which 
  seem to be related to the interesting  phase of the gravity.  The relation to the black hole for $p\to -\infty$ has been discussed in \cite{BrezinHikami5}.
  
  For two point function in the weak coupling region, we use the expression of (\ref{generalp}) with the value of $p= - 2$. Since the leading term in the exponent is $- c \frac{1}{y_i^4}$, we make a change of variable $ \frac{1}{y_i^4} = t_i$. Then the exponent becomes $- c t_i/(1+ s_i t_i)^2$. The two point function $u(s_1,s_2)$ is written as
  \ba\label{p-2}
  u(s_1,s_2)&=&  - \frac{1}{4} (\frac{s_2}{s_1})^{\frac{1}{2}} \oint \frac{dt_1 dt_2}{(2i\pi)^2} t_1^{-\frac{1}{2}} t_2^{-\frac{3}{2}} (1 + s_1 t_1)(1+ s_2 t_2) e^{-c\sum  \frac{t_i}{(1+ s_i t_i)^2}}
  \nonumber\\
  &\times& \frac{1}{1-g}
 \ea
  where $g$ is
  \be
  g = 2 s_1 t_1 + 2(\frac{s_2 t_1}{s_1 t_2})^{\frac{1}{2}}(1- s_2 t_2) - [s_1t_1 + (\frac{s_2 t_1}{s_1 t_2})^{\frac{1}{2}}(1- s_2 t_2) ]^2 - \frac{4 t_1}{s_1} (s_1+ s_2)^2
  \ee
  By the expansion of $1/(1-g) = \sum g^n$, we evaluate the terms of order $s_2^{n_2} s_1^{n_1}$.
  The selection rulein (\ref{RR})  becomes for two marked points of $p=-2$,
  \be
  g + 1 + \frac{1}{2}(j_1+ j_2) = n_1 + n_2
  \ee
  
 From (\ref{p-2}), we find the terms in the lower orders similar to (2.1)-(2.5) in the previous section. For $g=1$, $j_1=j_2=0$, we obtain the term $s_1^{\frac{1}{2}}s_2^{\frac{1}{2}}$ as
 \ba
 u(s_1,s_2) &=& c^{-1} s_1^{\frac{1}{2}} s_2^{\frac{1}{2}}  \int  \frac{dt_1 st_2}{(2 i \pi)^2} \frac{1}{2 t_2^{\frac{3}{2}}} ( t_1^{\frac{1}{2}} - t_1^{\frac{3}{2}})
 e^{- t_1 - t_2}\nonumber\\
 &=& - \frac{1}{16 \pi} s_1^{\frac{1}{2}}s_2^{\frac{1}{2}}
 \ea
  For the term of order $s_1 s_2$, which is $g=2$, $j_1=j_2= -1$,
  \ba
  u(s_1,s_2) &=& c^{-2} s_1 s_2 \int \frac{dt_1 dt_2}{(2 i \pi)^2} e^{- t_1 -t_2} \frac{1}{t_2^2} ( - 7 t_1^2 + 13 t_1^3 - t_1^4) \nonumber\\
  &=& - \frac{5}{64 \pi} s_1 s_2
  \ea
  where the integral of $t_2$ is a residue evaluation, and the integral $t_1$ is the exponential integral for $ 0 <t_1< \infty$. The coefficient $c$ is 
  $c = 4^{-p}= 4^2$.
  
  The terms $s_1^{n_1}s_2^{n_2}$ ($n_1,n_2$: integers) belong to Ramond sector ($j_1=j_2= -1$), and they appear in a pair. The terms of 
  $s_1^{n_1+ \frac{1}{2}} s_2^{n_2+ \frac{1}{2}}$ belong to Neveu-Schwarz sectors  with the spin components $j_1= j_2 = 0$
  in $u(s_1,s_2) \sim s_1^{n_1- \frac{1+ j_1}{2}}s_2^{n_2- \frac{1+ j_2}{2}}$.

    \vskip 2mm
        
     \vskip 2mm
    {\bf{7-4: Strong coupling of  $p= - 3$}} 
    \vskip 2mm
     For $p=-3$ case, the strong coupling expansion of the inverse of $s$ is
    \be
    u(s) = \frac{i}{2}\oint \frac{dy}{2i \pi} (y + \frac{1}{y^3}) e^{\frac{c}{s^2} \frac{y^6(y^4-1)}{(y^4+1)^4}}
    \ee
This large  $s$ expansion is given by the residue at $y = (-1)^{1/4}$,
    \ba\label{p=-3strong}
    &&u(s) = \frac{1}{s^4} + \frac{3}{8 s^8} + \frac{13}{2^6\cdot 3 s^{12}} + \frac{17\cdot 19}{2^{10}\cdot 3^2\cdot 5s^{16}} + \frac{23}{2^{14}\cdot 3 s^{20}}\nonumber\\
    && + \frac{29\cdot 31}{2^{17}\cdot 3^2\cdot 5\cdot 7 s^{24}} + \frac{29\cdot 31\cdot 37}{2^{20} \cdot 3^5\cdot 5^2\cdot 7 s^{28}} + 
    \frac{37\cdot 41\cdot 43}{2^{25}\cdot 3^4\cdot 5^2\cdot 7^2 s^{32}}\nonumber\\
    &&+ \frac{ 37\cdot 41\cdot 43\cdot 47}{2^{30}\cdot 3^6\cdot 5^3\cdot 7\cdot 11 s^{36}}+\frac{41\cdot 43\cdot 47\cdot 53}{ 2^{33} 3^8 5^3\cdot 7 \cdot 13 s^{40}}
   + \frac{47\cdot 53\cdot 59\cdot 61}{2^{36}3^7 5^4 \cdot 7\cdot 11\cdot 13 s^{44}} \nonumber\\
   &&+ \cdots
    \ea
    with $c= 2^6$. 
    These terms of order of $\frac{1}{s^{2k}}$ is expressed as
    \ba
    u(s) &=& \sum_{n=1}^\infty \frac{( 6 n-5)!!}{( 4 n-1)!! ( 2n-3)!! (n-1)! (n-1)!} \frac{3}{ n 2^{2n-2} } \frac{1}{s^{4 n}}\nonumber\\
    &=& \sum_{n=0}^\infty \frac{(6n+1)! (2 n+1)}{(4n+4)! ( 3n)! n!} \frac{8}{4^n s^{4n+4}}
    \ea

    With a logarithmic potential, $p=-3$ one point function $U(s)$ becomes
    \be
    U(s) = \oint \frac{du}{2i \pi}  e^{\frac{c}{s^2}( \frac{1}{(u+1)^2} - \frac{1}{(u-1)^2})} (\frac{u+1}{u-1})^N
     \ee
     Similar to $p=-2$ case, we make a change of variables of (\ref{changeuz}). Then we have
     \ba\label{p=-3stronglog}
     U(s) &=& - 2 \int dz e^{- N z} \frac{e^{-z}}{(1- e^{-z})^2} e^{-\frac{c}{4s^2} e^{2z} (1- e^{-z})^3 (1+ e^{-z})}\nonumber\\
     &=& -2 \int dz \frac{e^{-z}}{(1- e^{-z})^2} e^{\frac{c}{4 s^2} [ -2 e^{-z} + e^{-2 z} - e^{2 z} + 2 e^{z} ]} e^{-N z}
     \ea
     This gives a strong coupling expansion. 
     The term of order $\frac{c}{s^2}$ is 
     $$- (\frac{c}{s^2}) \frac{1}{N^2-1}$$
     The term of order $\frac{c^2}{s^4}$ is
     \ba
     &&- (\frac{c}{4 s^2})^2 \int dz e^{-Nz}e^{3z} (1- e^{-z})^4 (1+ e^{-z})^2\nonumber\\
     &&= - (\frac{c}{4 s^2})^2 \frac{48 ( 2 N^2-3)}{N (N^2-1)(N^2-4)(N^2-9)}
     \ea
     The order of $\frac{c^3}{s^6}$ in $U(s)$ for $p=-3$ becomes
     $$
     \frac{8965}{4}(\frac{c^3}{s^6}) \frac{N^2- 7}{(N^2-1)(N^2-4)(N^2-9)(N^2-16)(N^2-25)}
     $$
     The order of $\frac{c^4}{s^8}$ becomes
     $$
     - 9450 \frac{c^4}{s^8} \frac{2 N^4 - 38 N^2 + 63}{N(N^2-1)(N^2-4)(N^2-9)(N^2-16)(N^2-25)(N^2-36)(N^2-49)}
     $$
     The denominator is same as $p=-2$ case, but the numerator is different from $p=-2$ case. There is no term of order $\frac{1}{s^{2n+1}}$. The terms of order
     $\frac{1}{s^{4n}}$ have a factor $\frac{1}{N}$ and there is no such factor for other terms. When $N\to 0$, $U(s)$   becomes in the power series of $\frac{1}{s^{4n}}$. 
     Indeed the expansion of (\ref{p=-3strong}) agrees with the expansion of (\ref{p=-3stronglog}), and the coefficient is consistent with $c=2$.
    \vskip 2mm
    
          
          
                      \vskip 3mm
               \section{Summary and discussions}
    \vskip 3mm
    In this article, we have extended the results of previous articles I and II \cite{BrezinHikami1,BrezinHikami2} to $D_l$ type and to the multiple correlation functions of $p$ spin curve,
    specially for the non-positive integer values of $p$.
    
    The agreement with the  
     values, evaluated by the
    Gelfand Dikii equations \cite{LiuXu,LiuVakilXu}, has been shown in the Laurent expansions of $y$ variable.

    The intersection number of one marked point was examined for the large $p$ and large genus $g$. The analysis reveals interesting relations between the intersection numbers and the number theory through Bernoulli numbers. The intersection numbers are shown to be expressed as Bernoulli numbers ( for $p\to \infty$) multiplied a polynomial of $p$. 
    Therefore, the denominator of the intersection numbers have common values of the denominator of Bernoulli numbers. For $p=\frac{1}{2}$ spin case (fermionic), this denominator is cancelled by the numerator, and the intersection numbers are simply expressed as (\ref{onepoint1/2}).
    
      This integral representation enables us to continue the integer $p$ to the non-positive integer $p$, like $p= \frac{1}{2}$, $p=-\frac{1}{2}$ and $p=-2$ for which we 
      evaluated $n$ point functions explicitly as discussed in I and II \cite{BrezinHikami1,BrezinHikami2}.
      
      Since the central charge is given by $C= 2 - \frac{6}{p}$, the case $p= -\frac{1}{2}$ corresponds to  $C= 14$ for instance.  Such extension of the central charge $C$ ($C > 1$) is interesting from the view point of  conformal field theory, since  it goes over a barrier at $C=1$.
                The CFT for $C > 1$ has attracted interest for the case $1< C < 26$, which is related to 2d gravity coupled to matter field of the central charge $C$ \cite{BrezinHikami15}, and in this region, the behavior like a branched polymer is expected. The behavior  of quantum Liouville theory is discussed recently in the probabilistic approach \cite{Gwynne}. The conformal field theory of $p$ spin curves in the area of $1 < C$, $p < 0$ is interesting with respect to gauge theory and quantum gravity in higher dimensions and further studies are desirable.
          
    The $D_l$ type singularity has been investigated, under the new representation of the contour integrals  in the variables of $y_i$ for $m$-point correlation functions. The difference between $A_l$ and $D_l$ types is characterized by the factor, which appears as  the different measure  of the contour integrals. 
    We have shown that  $D_l$ type is obtained by the logarithmic
    term, which is expected from the supersymmetric random matrices \cite{BrezinHikami6}. 
     
  \vskip 3mm
  {\bf{Acknowledgement}}
  \vskip 2mm
  The author thank Edouard Br\'ezin for the encouragement and joint works, which appeared as I, II \cite{BrezinHikami1,BrezinHikami2}. He also thank Andreani Petrou
  for the discussions of knot polynomials.  This work is  supported by JSPS Kakenhi 19H01813.

  
 \end{document}